\documentclass[journal]{IEEEtran}
\usepackage{amsmath}
\usepackage{amsthm}
\usepackage{amssymb}
\usepackage{graphicx}
\usepackage{algorithm}
\usepackage{algorithmic}
\usepackage{cite}
\usepackage{color}
\usepackage{balance}
\newtheorem{theorem}{Theorem}
\newtheorem{lemma}{Lemma}

\ifCLASSINFOpdf
\else
\fi

\begin{document}
\title{Hyperbolic Frequency Multicarrier Modulation for Wideband Linear Time-Varying Channels}

\author{
	Xuehan~Wang,~\IEEEmembership{Graduate~Student~Member,~IEEE},
	Jinhong~Yuan,~\IEEEmembership{Fellow,~IEEE}, 
	Jintao~Wang,~\IEEEmembership{Senior~Member,~IEEE},
	Zhi~Sun,~\IEEEmembership{Senior~Member,~IEEE}, and
	JinXing,~Hao
	\thanks{This work was supported in part by the National Natural Science Foundation of China under Grants 624B2079 and 62271284. \textit{(Corresponding author: Jintao Wang)}.\par 
		Xuehan Wang, Jintao Wang and Zhi Sun are with the Department of Electronic Engineering, Tsinghua University, Beijing 100084, China, and also with the State Key laboratory of Space Network and Communications, Tsinghua University, Beijing 100084, China (e-mail: wang-xh21@mails.tsinghua.edu.cn;
		wangjintao@tsinghua.edu.cn; zhisun@ieee.org).\par 
		Jinhong Yuan is with the School of Electrical Engineering and Telecommunications, University of New South Wales, Sydney, NSW 2052, Australia (e-mail: j.yuan@unsw.edu.au).\par 
		Jinxing Hao is with the School of Accounting, Southwestern University of Finance and Economics, Chengdu, China (haojx@swufe.edu.cn).
	}
}

\maketitle
\begin{abstract} 
Numerous multicarrier modulation schemes have been proposed recently to enhance the performance in narrowband doubly dispersive channels for emerging high-mobility applications. However, the ultra-reliable modulation framework in wideband linear time-varying (LTV) channels remains an open problem, where the time dilations and contractions brought by the high mobility cannot be ignored for the baseband signal to obtain the constant Doppler shift across the whole transmission band. To solve this problem, we propose the hyperbolic frequency multicarrier (HFMC) waveform in this paper based on the inspiration from affine frequency division multiplexing (AFDM) modulation, where the delay and Doppler shift are absorbed into a 1D shift in the affine domain to provide a compact characterization of doubly dispersive discrete-time channels. By adopting the passband representation of wideband LTV channels and hyperbolic frequency modulated (HFM) signals, we reveal that the Doppler scaling factor brought by the relative mobility can be absorbed into an equivalent delay. The basic principle of HFMC modulation is established by investigating the approximate orthogonality among HFMC subcarriers, which are generated from a basic HFM signal by utilizing uniformly spaced equivalent delay. The spectrum of HFMC subcarriers is also analyzed to evaluate the system capacity, where the overlapping nature in the frequency domain can be observed. The input-output characterization in wideband LTV channels is then executed to confirm the 1D integration of time delay and Doppler scaling factor for each path, which demonstrates the ability to exploit potential multipath diversity. The parameter optimization based on the input-output relation and spectrum analysis is finally developed to balance the efficiency and reliability. Numerical results demonstrate the excellent bit error rate (BER) performance of the proposed HFMC waveform in wideband LTV channels.\par 
\end{abstract}
\begin{IEEEkeywords}
	Hyperbolic frequency multicarrier (HFMC) modulation, wideband linear time-varying (LTV) channels, multipath spread, Doppler scaling factor
\end{IEEEkeywords}
\IEEEpeerreviewmaketitle
\section{Introduction}
\label{sec_intro}
Multicarrier modulation has become one of the most essential enabling technologies in current wireless communication systems due to the high spectral efficiency and low-complexity equalization designs in time dispersive channels \cite{SC_OFDM_comparison,OTFS_tutorial,ref_multicarrier,satellite_shi}. However, the next-generation wireless networks are envisioned to support a wide range of emerging high-mobility applications, e.g., non-terrestrial networks \cite{satellite_TSP,shi_TSP_scenario,DL_modulation_mobility_us,satellite_shi}, and underwater acoustic (UWA) communications between autonomous underwater vehicles \cite{UWA_RIS_intro,UWA_RIS2_intro,UWA_OFDM_CL_harbin,UWA_OTFS_he_xibei}. In these scenarios, the conventional orthogonal frequency division multiplexing (OFDM) modulation suffers severe reliability degradation \cite{interference_tutorial,OFDM_degration} due to the time-varying property, which leads to severe inter-carrier interference and precision loss for channel estimation. Therefore, novel waveforms are expected to confront the challenges brought by the next-generation high-mobility applications \cite{OTFS_tutorial,interference_tutorial}. \par 
To address this issue, extensive modulation schemes have been proposed in recent years to enhance communication reliability by fully exploiting the multipath diversity \cite{OTFS_propose,ref_OTFS,ref_ODDM,ODDM_tutorial,ref_AFDM,AFDM_diversity_JSAC,diversity_TSP_General} of wireless channels. The authors in \cite{OTFS_propose} first developed the orthogonal time frequency space (OTFS) modulation schemes for doubly dispersive channels, whose data detection can be efficiently realized by iterative interference cancellation \cite{ref_OTFS,ODDM_linear_iterative} considering the sparse delay-Doppler (DD) domain channels. However, conventional OTFS systems would suffer severe performance degradation in practical implementation because of the high out-of-band emission (OOBE) brought by the discontinuous waveform \cite{OTFS_OOBE,ODDM_IO_Tcom_24}. Inspired by the basic idea of DD domain modulation schemes, the orthogonal delay-Doppler division multiplexing (ODDM) modulation has been proposed in \cite{ref_ODDM,ODDM_tutorial}. By adopting the Nyquist filter-based pulse-train as the delay-Doppler plane orthogonal pulse (DDOP), fine orthogonality with respect to DD resolutions can be acquired with lower OOBE. The interleave frequency division multiplexing modulation was proposed in \cite{ref_IFDM_WCL} to reconstruct the dense equivalent channel matrix, which exploits the detection gain by matching the low-complexity memory approximate message passing (MAMP) algorithm. In the meantime, the affine frequency division multiplexing (AFDM) modulation was proposed in \cite{ref_AFDM} to fully utilize the time-frequency diversity by employing a chirp-based discrete-time subcarrier design, where the time delay and Doppler shift can be integrated into a 1D delay in the affine domain. Thanks to the 1D compact representation, full multipath diversity and lower pilot overhead can be acquired by setting chirp parameters appropriately \cite{ref_AFDM,AFDM_diversity_JSAC}.\par 
It is worth pointing out that most of the aforementioned work focused on the narrowband channel model operating in the baseband signal, where the relative mobility is embodied as the constant Doppler shift across the whole transmission band for each incident path. As a result, significant performance deterioration occurs when facing wideband scenarios, e.g., the UWA communications, where the impact of Doppler scaling factors cannot be ignored even though the system still operates within the stationary time \cite{chenyang_UWA_simu,OTFS_DSE_TVT}. For example, the bit error rate (BER) floor has been observed in \cite{OTFS_DSE_TWC,OTFS_DSE_TVT} due to the mismatch between the OTFS modulation and wideband linear time-varying (LTV) channels. The deployment of advanced receivers, such as deep learning-based ones \cite{OTFS_DSE_WCL}, can mitigate the problems of time dilations and contractions brought by the Doppler scaling factors. However, the promotion was still limited \cite{OTFS_DSE_TVT,OTFS_DSE_WCL,ref_ODSS}, which necessitates the innovation of novel waveforms considering the unique property of wideband LTV channels.\par  
To the best of the authors' knowledge, limited research has been conducted for modulation designs in wideband LTV channels. The authors in \cite{ref_ODSS} developed the orthogonal delay scale space (ODSS) modulation inspired by the OTFS signaling, which enables excellent reliability with single-tap linear equalizers. However, it revealed significantly reduced spectral efficiency due to the robust bi-orthogonal assumptions and impractical implementation since it only considered the scaling effect and ignored the Doppler shift in the baseband signal. Another type of waveform was developed based on the hyperbolic frequency-modulated (HFM) signal. The authors in \cite{ref_hyperbolic_reviewer} proposed the hyperbolic secant root-raised cosine pulses to reduce inter-symbol-interference (ISI) in faster-than-Nyquist (FTN) systems. HFM signaling has also been investigated in wideband LTV channels by utilizing the shift keying (SK) techniques in chirp spread spectrum modulation systems \cite{CSS_modulation}, where the inherent scale-invariance \cite{HFM_signal_procedding} of HFM signals was exploited \cite{HFM_FSK_1_ICASSP,HFM_FSK_2_ocean,HFM_commu_MU,HFM_commu_SU}. However, the system capacity was also significantly low because of the employment of SK, where only $\log_{2}M$ bits can be transmitted by $M$ wideband HFM subcarriers with extremely large space in the frequency domain. To mitigate this problem, the orthogonal hyperbolic frequency division multiplexing (OHFDM) modulation has been developed in \cite{OHFDM_mine} by loading data symbols in each subcarrier with equally spaced frequency modulation (FM) rates. Unfortunately, OHFDM only reveals performance superiority when the Doppler scaling factors of each path are nearly the same, which is not suitable for general physical scenarios. Therefore, ultra-reliable modulation designs remain an open problem for wideband LTV channels, where comparable spectral efficiency with conventional OFDM systems is desired like OTFS, ODDM, and AFDM schemes in narrowband scenarios. \par 
On the other hand, although the modulation strategies developed for narrowband doubly dispersive channels may not be suitable for wideband LTV channels, the basic principle behind them could still be helpful in motivating novel schemes when the time dilations and contractions should be seriously considered to characterize the high-mobility channels. For example, OTFS has inspired the development of ODSS modulation \cite{ref_ODSS}. In fact, the work in this paper gains insight from AFDM designs \cite{ref_AFDM}, even though the exact continuous-time waveform and variants for wideband LTV channels remain open problems. The high-level idea of AFDM signaling is the 1D compact representation of doubly dispersive channels, where the potential multipath diversity can be maintained with appropriate chirp parameters. Since the Doppler scaling factor can be absorbed into an equivalent delay for HFM signals, a similar phenomenon might be expected for wideband LTV scenarios by considering the passband representation. Motivated by this idea, we propose the hyperbolic frequency multicarrier (HFMC) modulation in this paper. By investigating the orthogonality with respect to equivalent delay resolutions based on HFM signaling, the trade-off between spectral efficiency and potential diversity can be acquired for HFMC systems. Numerical results verify the excellent reliability of the proposed HFMC modulation scheme under appropriate parameter settings. To be more specific, the contributions of this paper can be summarized as follows. \par 
\begin{itemize}
	\item Based on the passband representation of wideband LTV channels and general multicarrier systems, we find that the Doppler scaling factor can be absorbed into the equivalent delay and complex gain for HFM signals. Therefore, the basic form for HFMC subcarriers is derived by appending an equally spaced equivalent delay. The approximate orthogonality with respect to the equivalent delay space (resolution) is theoretically analyzed, which establishes the principle of HFMC signaling.
	\item The spectrum occupation of near-orthogonal HFMC subcarriers is investigated by utilizing the method of stationary phase \cite{stationry_phase_method}. Different subcarriers reveal the overlapping nature in the frequency domain like OFDM, which is beneficial for increasing the spectral efficiency.
	\item The input-output relation for HFMC modulation in wideband LTV channels with both delay and Doppler scaling spread is characterized, where the approximate 1D shift property for each path can be confirmed.
	\item The parameter selection approach is developed for HFMC modulation by considering the trade-off between spectral efficiency and bandwidth of the equivalent channel matrix, which determines the potential diversity order similar to ODDM and AFDM systems. Numerical results finally verify the excellent BER performance of the proposed HFMC modulation scheme under conventional linear minimum mean square error (LMMSE) equalization.
\end{itemize}\par
It is worth noting that existing chirp-based multicarrier waveforms, such as AFDM \cite{ref_AFDM}, ODSS \cite{ref_ODSS}, and OHFDM \cite{OHFDM_mine}, have indeed exploited chirp-like pulses to enhance the performance, and their contributions have inspired the work presented in this paper. However, AFDM is fundamentally designed under the narrowband assumption with Doppler shift rather than scaling, rendering it less effective in wideband scenarios. ODSS, on the other hand, suffers from the unrealistic robust bi-orthogonal constraints and limited practical applicability, as its design accounts for time scaling effects while neglecting the Doppler shift component in the baseband signal. Similarly, OHFDM exhibits performance degradation in the presence of Doppler scaling spread because it is designed under path-invariant Doppler assumptions. In contrast, by exploiting the equivalent delay spread property inherent in HFM signaling, the proposed HFMC waveform is specifically designed to operate reliably under both delay and Doppler scaling spread, while preserving the desirable 1D compact representation of wideband LTV channels like AFDM in narrowband scenarios. Accordingly, this work fills the research gap by enabling the effective deployment of chirp-based multicarrier waveforms in wideband LTV channels.\par 
The rest of this paper is organized as follows. The passband representation of multipath wideband LTV channels and multicarrier modulation is first briefly reviewed in Section \ref{sec_system}. The basic principle and spectrum analysis of the proposed HFMC modulation are developed in Section \ref{sec_HFMC}. The input-output characterization in wideband LTV channels and parameter selection issues are presented in Section \ref{sec_HFMC_channel}. The numerical results are illustrated in Section \ref{sec_simu}. Finally, the conclusions are briefly drawn in Section \ref{sec_conclusion}.\par
\textit{Notations}: $\mathbf{A}$, $\mathbf{a}$, and $a$ denote the matrix, column vector, and scalar, respectively. $\mathbf{A}_{mn}$ represents the $(m,n)$-th component of $\mathbf{A}$. $(\cdot)^{*}$ and $\lceil\cdot\rceil$ stand for the conjugate and ceiling operation, respectively.\par 
\section{System Model}
\label{sec_system}
As illustrated in \cite{delay_scale_TSP,ref_VBMC,TSP_zhou_channel_simu_1,UWA_OTFS_ICSP,OTFS_DSE_TWC,OTFS_DSE_TVT}, the baseband characterization of wideband LTV channels is more complicated because the relative mobility will be embodied by both frequency shift and time scaling. Therefore, we adopt the more simplified passband representation and design the modulation accordingly in this paper. In this section, we first present the multipath channel model in wideband LTV systems with both delay and Doppler scaling spread. The passband representation of multicarrier modulation is then depicted.
\subsection{Passband Multipath Wideband LTV Channels}
\label{subsec_channel}
Following the illustrations in \cite{delay_scale_TSP,UWA_OFDM_dl_samek,TSP_zhou_channel_simu_1,OTFS_DSE_TVT,UWA_OFDM_CL_harbin,OFDM_passband_TVT_harbin,OHFDM_mine}, the multipath channels with delay and Doppler scaling spread are adopted to characterize the input-output relation at the waveform level. Let $s(t)$ denote the modulated passband signal, which is sent from the transmitter to the receiver via $P$ incident paths. The received passband signal at the receiver can then be derived as
\begin{equation}
	r(t)=\sum_{i=1}^{P}h_{i}(t)s\Big(t-\tau_{i}(t)\Big)+w(t),
	\label{passband_IO_waveform_ini}
\end{equation} 
where $h_{i}(t)$ and $\tau_{i}(t)$ denote the amplitude and time-varying delay associated with the $i$-th path, respectively, and $w(t)$ represents the additive white Gaussian noise. Within the stationary time \cite{delay_scale_TSP}, the path amplitude is constant as $h_{i}(t)\approx h_{i}$ while the time-varying delay can be approximated by $\tau_{i}(t)\approx\tau_{i}-a_{i}t$, where $a_{i}$ represents the Doppler scaling factor of the $i$-th path. As a result, \eqref{passband_IO_waveform_ini} can be rewritten as
\begin{equation}
	r(t)=\sum_{i=1}^{P}h_{i}s\Big((1+a_{i})t-\tau_{i}\Big)+w(t).
	\label{passband_IO_waveform}
\end{equation}\par 
From the illustrations and experimental validation in \cite{OTFS_DSE_TVT,TSP_zhou_channel_simu_1,UWA_MIMO_channel_TSP,ref_LTVchannel_book}, the Doppler scaling factor can be derived as $a_{i}=\frac{v_{i}}{c}$, where $c$ and $v_{i}$ represent the propagation speed of the communication medium in the specific transmission environment and the velocity with which the $i$-th path length is decreasing, respectively. For example, the typical value of $c$ is around $1500$ m/s for UWA communications, which leads to the scaling factor of more than $3\times10^{-3}$ when the maximum relative mobility is more than $10$ kn. It is significantly different from conventional radio-frequency (RF) communications where the scaling factors can be ignored in the baseband signal by considering $|a_{i}|<10^{-6}$ to derive the widely-used approximation of Doppler shift \cite{ref_ODDM,ref_OTFS,ref_AFDM}. Therefore, we propose to process the passband signal directly to provide a more simplified characterization in \eqref{passband_IO_waveform} in this paper, where the Doppler shift will not be involved. Finally, we assume $\tau_{i}<\tau_{\max}$ and $|a_{i}|\leq a_{\max}$ to determine the delay and Doppler scaling spread, where we have $a_{\max}=\frac{v_{\max}}{c}$ with $v_{\max}$ denoting the maximum relative mobility speed. Finally, it is essential to note that the Doppler scaling factors associated with individual propagation paths are generally distinct. The proposed HFMC modulation is capable of handling this general scenario, in which it effectively transforms the 2D delay and Doppler scaling spread into the 1D equivalent delay spread.\par 
\subsection{Multicarrier Modulation with Passband Representation}
\label{subsec_multicarrier}
In this subsection, the continuous-time passband representation of general multicarrier modulation schemes is presented like \cite{OHFDM_mine,ODDM_tutorial,OFDM_passband_TVT_harbin,UWA_OFDM_dl_samek,UWA_OFDM_begin,ref_multicarrier}. Let $T$ and $M$ denote the time duration of each multicarrier symbol and the number of subcarriers within a transmit multicarrier symbol, the transmit waveform $s(t)$ can be derived by
\begin{equation}
	\label{MC_basicform_general}
	s(t)=\sum_{m=0}^{M-1}x_{m}\varphi_{m}(t),~-T_{p}\leq t\leq(1+a_{\max})T,
\end{equation}
where $\varphi_{m}(t)$ and $x_{m}$ denote the $m$-th subcarrier and the data symbol loaded by this subcarrier, respectively. To avoid the interference between adjacent multicarrier symbols, the prefix and suffix should be added, whose time duration can be denoted as $T_{p}>\tau_{\max}$ and $a_{\max}T$, respectively. Considering the basic principle of passband representation like \cite{ODDM_tutorial,OFDM_passband_TVT_harbin,UWA_OFDM_dl_samek,UWA_OFDM_begin}, the spectrum of $\varphi_{m}(t)$ should be approximately\footnote{Considering the finite time span, the spectrum cannot be strictly band-limited, which leads to out-of-band emission (OOBE) issues and will be potentially embodied in our future work.} limited within $(f_{c}-\frac{B}{2},f_{c}+\frac{B}{2})$, where $f_{c}$ and $B$ denote the center frequency and bandwidth occupation, respectively.\par 
Similar to modulation processing, demodulation is performed at the receiver by the corresponding matched filters or correlators, which can be derived as 
\begin{equation}
	\begin{aligned}
		&y_{n}=\int_{0}^{T}\phi_{n}^{*}(t)r(t)dt\\
		&=\int_{0}^{T}\phi_{n}^{*}(t)\bigg(\sum_{i=1}^{P}h_{i}s\Big((1+a_{i})t-\tau_{i}\Big)\bigg)dt+w_{n},
	\end{aligned}
\end{equation} 
where $\phi_{n}(-t)$ and $w_{n}$ represent the matched filter and equivalent additive noise sample associated with the $n$-th subcarrier, respectively. In this paper, we assume $\phi_{n}(t)=\varphi_{n}(t)$ to simplify the analysis while they can be different in general to obtain more performance gains \cite{ref_multicarrier,ref_modulation_book}. Taking conventional cyclic prefix (CP)-OFDM systems as an example in \cite{UWA_OFDM_dl_samek,UWA_OFDM_begin}, $\varphi_{m}(t)$ can be derived as
\begin{equation}
	\varphi_{m}(t)=\frac{1}{\sqrt{T}}e^{j2\pi \left(f_{c}-\frac{B}{2}+\frac{m}{T}\right)t}.
\end{equation}\par 
At last, the extracted components $y_{n}$ are passed to the equalizer to recover the data symbols $x_{m}$ as $\hat{x}_{m}$, which can be utilized to attain transmit data bits.\par 

\section{Hyperbolic Frequency Multicarrier Modulation}
\label{sec_HFMC}
In AFDM systems \cite{ref_AFDM}, the chirp waveform enables the integration of time delay and Doppler shift into a 1D shift in the affine domain, where the Doppler shift is effectively translated into an additional equivalent delay. With appropriate chirp parameters, this design can achieve both potential diversity gain and compact representation of narrowband LTV channels. Inspired by this principle, we propose HFMC modulation in this section for passband communications in wideband LTV systems, where AFDM becomes ineffective because of the extreme Doppler scaling spread, resulting in time dilations and contractions. To address this issue, the HFM signal is adopted in this section, whose logarithmic phases absorb the Doppler scaling factor and convert it into an additional equivalent delay. Thus, while AFDM operates in the baseband affine domain under the narrowband assumption, HFMC extends this concept to the wideband scenarios by converting the Doppler scaling effect into a delay-like structure. In this section, we first briefly review the basic concepts of HFM signal and illustrate why the Doppler scaling factor can be absorbed into an equivalent delay. The subcarriers of HFMC modulation are then derived considering the approximate orthogonality. Finally, the spectrum analysis is developed to measure the spectral efficiency and determine the conditions for parameter selection.  \par 
\subsection{HFM Signal}
\label{subsec_HFS}
Since the scaling factor can be absorbed into an equivalent delay \cite{HFM_commu_MU,HFM_signal_procedding} in HFM signals, we adopt the HFM signals to provide convenience for signal processing. Similar to \cite{HFM_commu_MU,HFM_signal_procedding,HFM_FSK_2_ocean}, a general HFM signal can be derived as
\begin{equation}
	g(t)=\frac{1}{1+\frac{t}{T_{0}}}e^{j2\pi K\ln\left(1+\frac{t}{T_{0}}\right)},
	\label{HFM_signal_basic}
\end{equation}
where $K>0$ and $T_{0}>0$ are the FM rate and reference time, respectively. In literature, HFM signals are also referred to as linear period modulation (LPM) signals since the instantaneous period varies linearly with the time increasing, where $\frac{T_{0}}{K}$ denotes the instantaneous period at $t=0$. Considering the time dilation or contraction brought by the Doppler scaling factor $a_{i}$, the received signal can be derived as
\begin{equation}
	\label{scaling_absorb}
	\begin{aligned}
		&g\big((1+a_{i})t\big)=\frac{1}{1+\frac{(1+a_{i})t}{T_{0}}}e^{j2\pi K\ln\left(1+\frac{(1+a_{i})t}{T_{0}}\right)}\\
		&=\frac{1}{(1+a_{i})\big(1+\frac{t-\frac{a_{i}T_{0}}{1+a_{i}}}{T_{0}}\big)}e^{j2\pi K\ln\left((1+a_{i})\big(1+\frac{t-\frac{a_{i}T_{0}}{1+a_{i}}}{T_{0}}\big)\right)}\\
		&=\frac{e^{j2\pi K\ln(1+a_{i})}}{1+a_{i}}g(t-\dot{t}_{i}),
	\end{aligned}
\end{equation}
where the equivalent delay is $\dot{t}_{i}=\frac{a_{i}}{1+a_{i}}T_{0}$. Eq. \eqref{scaling_absorb} indicates that by adopting HFM signals $g(t)$ as the basic transmission waveforms, the Doppler scaling operation on the signal $g((1+a_{i})t)$ can be represented by a delayed version of the transmit signal as $g(t-\dot{t}_{i})$, where the equivalent delay $\dot{t}_{i}$ is determined by the Doppler scaling factor. It can be observed that, similar to how an AFDM system can convert the Doppler shift of a baseband signal into an additional equivalent delay, the HFM signal can transform the Doppler scaling factor of a passband signal into an additional equivalent delay. Therefore, it is expected that by applying HFM signal-based modulation, a compact 1D representation analogous to that of AFDM in narrowband LTV channels can be achieved in wideband scenarios, while maintaining resolution for different channel paths to exploit potential multipath diversity gain. \par 
\begin{figure}
	\centering{\includegraphics[width=0.9\linewidth]{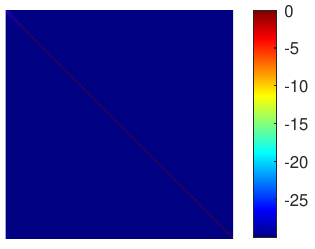}}
	\caption{The correlation (dB) among transmit subcarriers.}
	\label{Fig_orthogonality}	
\end{figure}
\subsection{HFMC Modulation}
\label{subsec_HFMC}
As shown in the aforementioned subsection, the Doppler scaling can be transferred into an equivalent time delay for HFM signals, which leads to the 1D spread after passing through the wideband LTV channels. Therefore, different subcarriers are expected to generate from the basic HFM signal in \eqref{HFM_signal_basic} with different equivalent time delays, which can help exploit the multipath diversity with both delay and Doppler scaling spread. To be more specific, the $m$-th subcarrier for the HFMC systems is derived from \eqref{HFM_signal_basic} with the equivalent delay $t_{m}$ as
\begin{equation}
	\label{subcarrier_form}
	\begin{aligned}
		\varphi_{m}(t)&=A_{m}g(t-t_{m})\\
		&=\frac{A_{m}}{1+\frac{t-t_{m}}{T_{0}}}e^{j2\pi K\ln\left(1+\frac{t-t_{m}}{T_{0}}\right)},
	\end{aligned}
\end{equation} 
where the time range is by $-T_{p}\leq{t}\leq{(1+a_{\max})T}$ and $0\leq{t}\leq{T}$ at the transmitter and receiver, respectively, and $A_{m}$ denotes the amplitude coefficient to normalize the power of the $m$-th subcarrier. Since all subcarriers occupy the whole time duration of a multicarrier symbol, it cannot be treated as a single-carrier system even though subcarriers are distinguished with different equivalent delays. In fact, the spectrum of each subcarrier also reveals a similar shift property, which is depicted in detail in the next subsection. \par 
Similar to the standard analysis of multicarrier systems \cite{ref_multicarrier,UWA_OFDM_DHT_passband,ref_modulation_book,ODDM_tutorial,UWA_OFDM_book,ref_LTVchannel_book}, the orthogonality among subcarriers should be guaranteed at the transmitter, which also leads to the fine resolution for the integrated equivalent time delay. However, it is almost impossible to ensure strict orthogonality at the transmitter considering the complicated form of HFMC subcarriers. Instead, we utilize the approximate orthogonality by selecting appropriate space for $t_{m}$. At first, the following theorem can be established to serve as the basis for orthogonality analysis.\par 
\begin{theorem}
	\rm 
	\label{th1_ideal_correlation}
	The cross-correlation between subcarriers can be derived as
	\begin{equation}
		\begin{aligned}
			&\int_{0}^{T}\varphi_{n}^{*}(t)\varphi_{m}(t)dt\\
			&=\frac{A_{m}A_{n}T_{0}^{2}Te^{j\pi K\theta_{nm}}}{(T_{0}+T-t_{n})(T_{0}-t_{m})}\\
			&\times\frac{\sin\pi{K}\ln\left(1+\frac{(t_{m}-t_{n})T}{(T_{0}+T-t_{n})(T_{0}-t_{m})}\right)}{\pi{K}\frac{(t_{m}-t_{n})T}{(T_{0}+T-t_{n})(T_{0}-t_{m})}}\\
			&\overset{(a)}{\approx}\frac{A_{m}A_{n}T_{0}^{2}T}{(T_{0}+T-t_{n})(T_{0}-t_{m})}\\
			&\times  e^{j\pi K\theta_{nm}}\text{sinc}\left(\frac{K(t_{n}-t_{m})T}{(T_{0}+T-t_{n})(T_{0}-t_{m})}\right)\\
			&\overset{(b)}{\approx}e^{j\pi K\theta_{nm}} \text{sinc}\left(\frac{K(t_{n}-t_{m})T}{(T_{0}+T)T_{0}}\right),
		\end{aligned}
		\label{ideal_correlation}
	\end{equation}
	where we have $\text{sinc}(x)=\frac{\sin \pi x}{\pi x}$ and $A_{m}=\sqrt{\frac{(T_{0}+T-t_{m})(T_{0}-t_{m})}{T_{0}^{2}T}}$ to normalize the transmit power for each subcarrier as in \eqref{subcarrier_form}. The subcarrier-dependent phase rotation $\theta_{nm}$ can be derived as
	\begin{equation}
		\theta_{nm}=\ln\left(\frac{(T_{0}-t_{m})(T_{0}+T-t_{m})}{(T_{0}-t_{n})(T_{0}+T-t_{n})}\right).
	\end{equation} 
	The approximation in \eqref{ideal_correlation}(a) is valid by assuming $\ln(1+x)\approx x$ with $x=\frac{(t_{m}-t_{n})T}{(T_{0}+T-t_{n})(T_{0}-t_{m})}$. This assumption is ensured when the system parameters satisfy $|t_{m}|\ll T_{0}$, which also leads to the simplified formulation in \eqref{ideal_correlation}(b). To characterize the approximation accuracy, an error threshold $\epsilon$ is introduced such that $|t_{m}|<\epsilon T_{0}$. Given that $|\ln(1+x)-x|<0.55x^{2}$ for $|x|<0.1$ and $T<T_{0}$, the linearization error of the logarithmic terms is bounded by $0.55\epsilon^{2}<6\times10^{-3}$, thereby guaranteeing the validity of the above analysis. As will be shown in the following illustrations, $\epsilon$ can be chosen sufficiently small, e.g., $\epsilon=0.034$ in the simulations, to ensure the validity of \eqref{ideal_correlation}.
	\begin{IEEEproof}
		The proof is completed in Appendix \ref{proof_th1_ideal_correlator}.
	\end{IEEEproof}
\end{theorem}
From the derivations in \eqref{ideal_correlation}, the equivalent delay resolution of $t_{r}=\frac{(T_{0}+T)T_{0}}{KT}$ can be acquired if $|t_{m}|\ll T_{0}$ since we have $\text{sinc}(x)=0$ for $x\in\mathbb{Z}\setminus{\{0\}}$. As a result, the equivalent delay for the $m$-th subcarrier can be selected as
\begin{equation}
	t_{m}=t_{0}+mt_{r}=t_{0}+m\frac{(T_{0}+T)T_{0}}{KT},
\end{equation}
which guarantees the approximate orthogonality among subcarriers. Fig. \ref{Fig_orthogonality} confirms the correctness of \textbf{Theorem \ref{th1_ideal_correlation}} by plotting the correlation among subcarriers under simulation parameters in Section \ref{sec_simu}, where the approximate orthogonality can be confirmed since only very slight leakage occurs at edge subcarriers.\par 
\begin{figure}
	\centering{\includegraphics[width=1\linewidth]{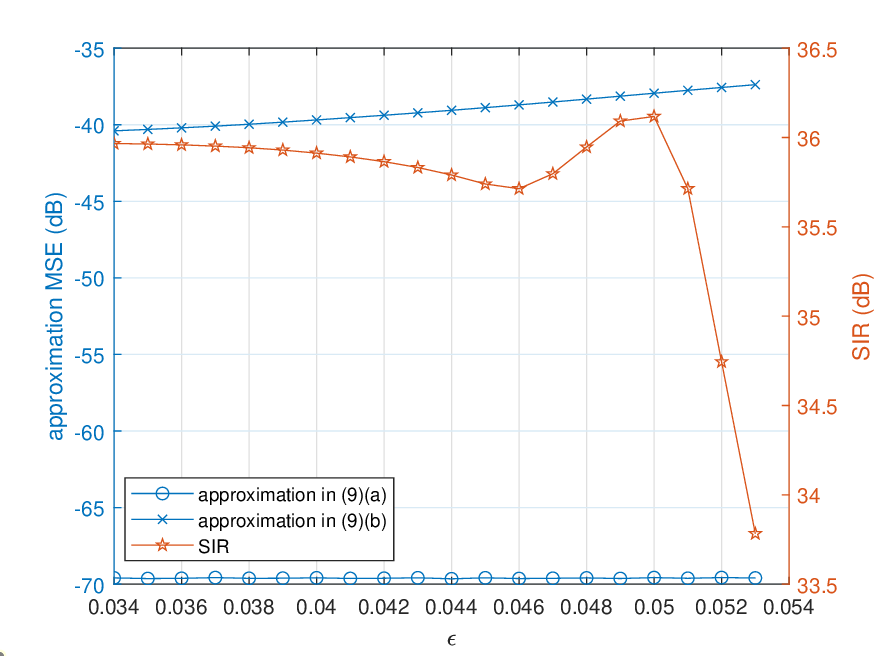}}
	\caption{SIR and approximation MSE evaluation for \textbf{Theorem \ref{th1_ideal_correlation}}.}
	\label{Fig_approMSE_th1}	
\end{figure}
The accuracy of the approximate orthogonality analysis in \textbf{Theorem \ref{th1_ideal_correlation}} is further quantified in Fig. \ref{Fig_approMSE_th1} by illustrating the signal-to-interference ratio (SIR) and approximation mean squared error (MSE) under simulation parameters in Table \ref{table_simu}. As observed in Fig. \ref{Fig_approMSE_th1}, the approximation MSE increases with the error threshold $\epsilon$, which aligns with the analysis in \textbf{Theorem \ref{th1_ideal_correlation}}. For $\epsilon<0.036$, the approximation MSE in $(9)(b)$ falls below $-40$ dB, ensuring its practical reliability. Meanwhile, the SIR exceeds $35.5$ dB for $\epsilon\leq0.05$, which further corroborates the validity of the approximate orthogonality. To guarantee the applicability, $\epsilon$ should be selected as small as possible, guiding the parameter selection in Section \ref{subsec_parameter}.\par 
Based on the above analysis, the matrix notation of the proposed HFMC modulation can also be derived by considering its sequence-level characterization. As in conventional assumptions \cite{ref_LTVchannel_book,UWA_OFDM_dl_samek,OHFDM_mine}, oversampling is required to address OOBE, i.e., the sampling frequency $F_{s}$ is set significantly higher than the signal bandwidth $B$. For example, the sampling frequency is given by $F_{s}=\gamma B$, where the oversampling factor $\gamma$ is an integer greater than one. Therefore, the number of time domain samples without considering the prefix and suffix can be derived as $K=\lceil{TF_{s}}\rceil$. At the transmitter, the time domain sequence $\mathbf{s}\in\mathbb{C}^{K\times 1}$ is obtained as
\begin{equation}
	\mathbf{s}=\mathbf{\Phi}\mathbf{x},
\end{equation}
where $\mathbf{\Phi}\in\mathbb{C}^{K\times M}$ is the discrete modulation matrix and $\mathbf{x}$ denotes the data vector. Specifically, the entries of $\mathbf{\Phi}$ are given by $\mathbf{\Phi}_{km} = \varphi_{m}(kT_{s})$, and $\mathbf{x}_{m}=x_{m}$ represents the $m$-th data symbol. After that, the transmitted sequence can be finalized by adding the prefix and suffix, which take the waveform continuity into consideration by following \eqref{MC_basicform_general} and \eqref{subcarrier_form} rather than the cyclic extension.\par 
At the receiver side, the time domain sequence is first attained after sampling and discarding the prefix and suffix, which leads to the time domain input-output relation as 
\begin{equation}
	\mathbf{r}=\mathbf{H}_{\text{T}}\mathbf{s}+\mathbf{w}_{\text{T}},
\end{equation}
where $\mathbf{H}_{\text{T}}\in\mathbb{C}^{K\times K}$ and $\mathbf{w}_{\text{T}}\in\mathbb{C}^{K\times1}$ represent the time domain channel matrix and additive white Gaussian noise samples, respectively. The matched filter-based demodulation is then employed as
\begin{equation}
	\label{matrix_IO_1}
	\mathbf{y}=\mathbf{\Phi}^{H}\mathbf{r}=\mathbf{Hx}+\mathbf{w},
\end{equation}
where the equivalent channel matrix and additive noise are respectively given by $\mathbf{H}=\mathbf{\Phi}^{H}\mathbf{H}_{\text{T}}\mathbf{\Phi}$ and $\mathbf{w}=\mathbf{\Phi}^{H}\mathbf{w}_{\text{T}}$. In view of the approximate orthogonality established in \textbf{Theorem \ref{th1_ideal_correlation}}, $\mathbf{w}$ can be treated as additive white Gaussian noise, while a detailed analysis of the equivalent channel matrix $\mathbf{H}$ is provided in Section \ref{subsec_IO}.\par 
Based on the above analysis, the implementation complexity of the proposed HFMC waveform is $\mathcal{O}(KM)$. This complexity is higher than that of conventional OFDM systems, which is $\mathcal{O}(K\log{M})$ due to the fast Fourier transform (FFT)-based implementation. The increased complexity arises from the more elaborate signal structure designed to achieve superior reliability and a more compact representation of wideband LTV channels. The proposed scheme also preserves the resolvability of individual multipath components according to the following analysis in Section \ref{subsec_IO}, thereby enabling the exploitation of potential multipath diversity gain. While the computational cost of the waveform itself is higher, it remains acceptable for practical parameter ranges, and the performance improvements presented in Section \ref{sec_simu} justify the value of the additional overhead. Our future work will focus on developing reduced-complexity implementations to further enhance the practicality of the proposed waveform.\par
\subsection{Spectrum Analysis}
\label{subsec_spectrum}
In this subsection, the spectrum of HFMC subcarriers is analyzed to determine the available range of parameters $t_{m}$ by utilizing the method of stationary phase \cite{stationry_phase_method}. We first investigate the spectrum of the transmit subcarriers, which is presented in the following theorem.
\begin{theorem}
	\label{th2_spectrum}
	\rm 
	If $KT^{2}\gg T_{0}^{2}+T_{0}T$, the spectrum of $\varphi_{m}(t)$ can be approximately derived as
	\begin{equation}
		\label{spectrum_subcarrier}
		\begin{aligned}
			\psi_{m}(f)&=\int_{-T_{p}}^{(1+a_{\max})T}\varphi_{m}(t)e^{-j2\pi ft} dt\\
			&\approx\frac{A_{m}T_{0}e^{-j\frac{\pi}{4}}}{\sqrt{K}}e^{j2\pi\theta(f)},~ f_{m}^{s}\leq f\leq f_{m}^{e}
		\end{aligned}
	\end{equation}
	where we have 
	\begin{equation}
		\label{spectrum_angle_subcarrier}
		\theta(f)=K\ln\left(\frac{K}{fT_{0}}\right)-K-f(t_{m}-T_{0}).
	\end{equation}
	$(f_{m}^{s},f_{m}^{e})$ denotes the approximate frequency span of the $m$-th subcarrier as
	\begin{equation}
		\begin{cases}
			f_{m}^{s}&=\frac{K}{T_{0}+(1+a_{\max})T-t_{m}}\\[2bp]
			f_{m}^{e}&=\frac{K}{T_{0}-T_{p}-t_{m}}
		\end{cases}.
	\end{equation}
	\begin{IEEEproof}
		The proof is completed in Appendix \ref{proof_th2_spectrum}.
	\end{IEEEproof}
\end{theorem}
Based on the deduction in \textbf{Theorem \ref{th2_spectrum}}, the amplitude spectrum of each subcarrier is almost flat within the frequency span. When $|t_{m}-t_{n}|<(1+a_{\max})T+T_{p}$, the frequency occupation of the $m$-th and $n$-th subcarriers would overlap\footnote{Based on the parameter analysis and numerical results in Fig. \ref{Fig_spectrum}, it is very possible to occur since a high resolution of the equivalent delay is always desired.}. On the other hand, all subcarriers are required to be band-limited within the interval $(f_{c}-\frac{B}{2},f_{c}+\frac{B}{2})$. Therefore, we have $f_{m}^{s}>f_{c}-\frac{B}{2}$ and $f_{m}^{e}<f_{c}+\frac{B}{2}$ for $\forall~m$, which can be translated into
\begin{equation}
	\label{tm_range}
	\begin{cases}
		t_{m}>T_{0}+(1+a_{\max})T-\frac{K}{f_{c}-\frac{B}{2}}\\
		t_{m}<T_{0}-T_{p}-\frac{K}{f_{c}+\frac{B}{2}}
	\end{cases}.
\end{equation}\par 
To guarantee that there are valid values for $t_{m}$ to satisfy \eqref{tm_range}, $K$ should be selected as
\begin{equation}
	\label{K_eq_1}
	K>\frac{(f_{c}-\frac{B}{2})(f_{c}+\frac{B}{2})}{B}\Big(T_{p}+(1+a_{\max})T\Big).
\end{equation}
Based on this derivation, to maximize the spectral efficiency while maintaining the approximate orthogonality by utilizing the results in \textbf{Theorem \ref{th1_ideal_correlation}}, $t_{m}$ can be selected as $t_{m}=t_{0}+mt_{r}$, where we have
\begin{equation}
	\label{t0_value}
	t_{0}=T_{0}+(1+a_{\max})T-\frac{K}{f_{c}-\frac{B}{2}}.
\end{equation}
Therefore, the maximum number of transmit subcarriers can be derived as
\begin{equation}
	\label{system_capacity}
	M=\left\lceil\frac{\frac{KB}{(f_{c}-\frac{B}{2})(f_{c}+\frac{B}{2})}-T_{p}-(1+a_{\max})T}{t_{r}}\right\rceil,
\end{equation}
where $\lceil\cdot\rceil$ denotes the ceiling operation. Eq. \eqref{system_capacity} is deduced by considering the available range in \eqref{tm_range}. Based on the derivations in \eqref{t0_value} and \eqref{system_capacity}, the system capacity can be characterized by the FM rate $K$ and reference time $T_{0}$. To select appropriate values for these parameters, the input-output characterization in UWA channels should be considered, which is illustrated in detail in Section \ref{subsec_parameter}.\par 
\section{HFMC over Wideband Time-Varying Channels}
\label{sec_HFMC_channel}
In this section, we first develop the input-output characterization for HFMC systems in multipath wideband LTV channels as shown in \eqref{passband_IO_waveform}, which reveals both the delay and Doppler scaling spread. Then we investigate the parameter optimization to obtain a flexible trade-off among the transmission efficiency, potential diversity, and receiver complexity by developing the selection framework.\par  
\subsection{Input-Output Characterization}
\label{subsec_IO}
We first consider the received component from the $i$-th path, which is characterized by the parameters $\tau_{i}$, $a_{i}$ and $h_{i}$. The channel-distorted continuous-time received signal for the $m$-th subcarrier can be derived as
\begin{equation}
	\label{eqdelay_double_channel}
	\begin{aligned}
		&h_{i}\varphi_{m}\Big((1+a_{i})t-\tau_{i}\Big)\\
		&=\frac{h_{i}A_{m}}{1+\frac{(1+a_{i})t-\tau_{i}-t_{m}}{T_{0}}}e^{j2\pi K\ln\left(1+\frac{(1+a_{i})t-\tau_{i}-t_{m}}{T_{0}}\right)}\\
		&=h_{i}\frac{e^{j2\pi\ln(1+a_{i})}}{1+a_{i}}A_{m}\times\frac{1}{1+\frac{t-\frac{\tau_{i}+t_{m}+a_{i}T_{0}}{1+a_{i}}}{T_{0}}}\\
		&\times e^{j2\pi K\ln\left(1+\frac{t-\frac{\tau_{i}+t_{m}+a_{i}T_{0}}{1+a_{i}}}{T_{0}}\right)}\\
		&=h_{i}^{\prime}A_{m}g(t-t_{m,i}^{\prime}),
	\end{aligned}
\end{equation}
where we have $h_{i}^{\prime}=h_{i}\frac{e^{j2\pi\ln(1+a_{i})}}{1+a_{i}}$ and $t_{m,i}^{\prime}=\frac{\tau_{i}+t_{m}+a_{i}T_{0}}{1+a_{i}}$. It is obvious that the Doppler scaling factor can be absorbed into the equivalent subcarrier-independent complex gain and subcarrier-dependent delay. Therefore, the following theorem can be established to illustrate the input-output relation for HFMC systems.\par 
\begin{theorem}
	\rm
	\label{th3_IO}
	Let $q_{nm}$ denote the channel coefficient between the $m$-th transmit subcarrier and $n$-th received subcarrier, i.e., $y_{n}=\sum_{m=0}^{M-1}q_{nm}x_{m}+w_{n}$. For ease of illustration, $q_{nm}$ can be divided as $q_{nm}=\sum_{i=1}^{P}h_{i}q_{nm}^{i}$ with $q_{nm}^{i}$ denoting the component from the $i$-th path. $q_{nm}^{i}$ can be derived as
	\begin{equation}
		\label{IO_onepath}
		\begin{aligned}
			&q_{nm}^{i}=\int_{0}^{T}\varphi_{n}^{*}(t)\varphi_{m}\Big((1+a_{i})t-\tau_{i}\Big)dt\\
			&=\frac{A_{m}A_{n}T_{0}^{2}Te^{j\pi K\theta_{nm}^{i}}}{(T_{0}+T-t_{n})(T_{0}-t_{m}-\tau_{i})}\\
			&\times  \text{sinc}\left(\frac{KT\big((1+a_{i})t_{n}-t_{m}-(\tau_{i}+a_{i}T_{0})\big)}{(T_{0}+T-t_{n})(T_{0}-t_{m}-\tau_{i})}\right)\\
			&\overset{(a)}{\approx}e^{j\pi K\theta_{nm}^{i}}\text{sinc}\left(\frac{KT\Big(t_{n}-t_{m}-(\tau_{i}+a_{i}T_{0})\Big)}{(T_{0}+T)T_{0}}\right),	
		\end{aligned}
	\end{equation}
	where the approximation in $(a)$ holds if we have $|t_{n}|\ll T_{0}$ and $|t_{m}+\tau_{i}|\ll T_{0}$. The subcarrier-dependent phase rotation $\theta_{nm}^{i}$ can be derived as 
	\begin{equation}
		\label{phase_onepath}
		\begin{aligned}
			&\theta_{nm}^{i}=\\
			&\ln\left(\frac{(T_{0}-t_{m}-\tau_{i})(T_{0}+(1+a_{i})T-t_{m}-\tau_{i})}{(T_{0}-t_{n})(T_{0}+T-t_{n})}\right).
		\end{aligned}		
	\end{equation}
	\begin{IEEEproof}
		The proof is completed in Appendix \ref{proof_th3_IO}.
	\end{IEEEproof}
\end{theorem}\par 
\par 
\begin{figure}
	\centering{\includegraphics[width=1\linewidth]{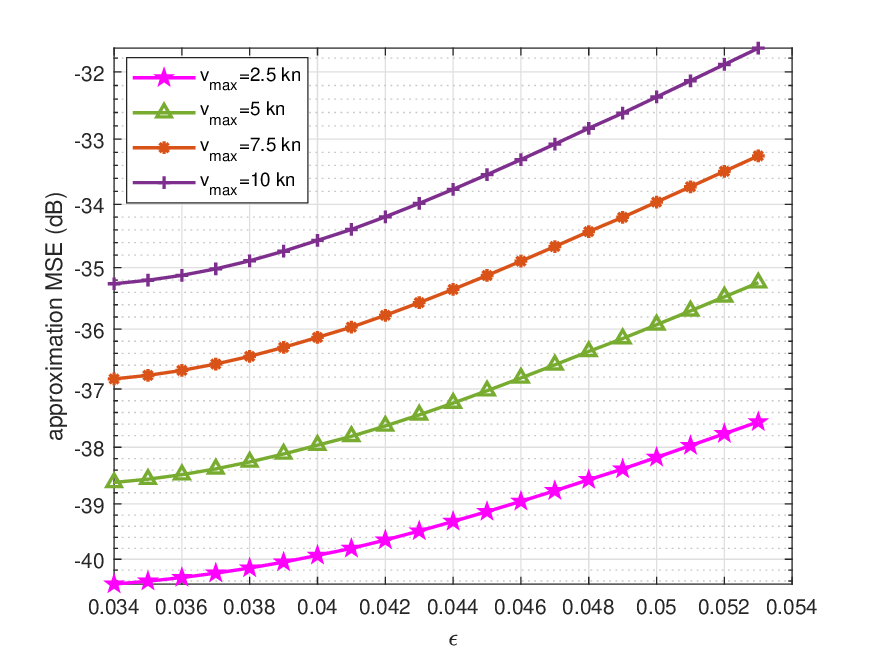}}
	\caption{Evaluation of approximation MSE in \textbf{Theorem \ref{th3_IO}}.}
	\label{Fig_approMSE_th3}	
\end{figure}
Fig. \ref{Fig_approMSE_th3} evaluates the accuracy of the approximation in \textbf{Theorem \ref{th3_IO}} by plotting the approximation MSE against the error threshold $\epsilon$ for different values of the maximum mobility velocity $v_{\max}$. As demonstrated, the MSE increases monotonically with both $\epsilon$ and $v_{max}$. Even with the highest considered mobility of $v_{max} = 10$ kn, the MSE remains below $-35$ dB for $\epsilon \leq 0.036$, which confirms the practical reliability of the approximation under the considered parameter settings.\par
\begin{figure}
	\centering{\includegraphics[width=0.9\linewidth]{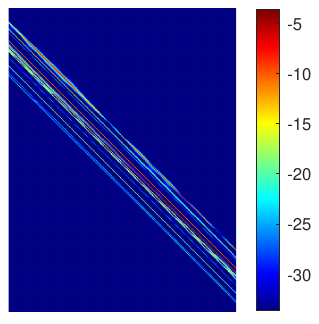}}
	\caption{Equivalent channel matrix (dB) of HFMC systems.}
	\label{Fig_eqChannel}	
\end{figure}

 
\textbf{Theorem \ref{th3_IO}} also confirms that HFMC extends the characterization of AFDM in narrowband LTV channels to wideband LTV channels. The proposed HFMC transforms a wideband LTV channel with 2D spread into a representation with a 1D equivalent delay spread. Moreover, based on the differing equivalent delays of various channel paths, potential multipath diversity gain can be exploited. To be more specific, it can be observed that the time delay and Doppler scaling factor of each path are integrated into an equivalent delay as $\tau_{i}+a_{i}T_{0}$, which also provides the possibility of low-overhead channel estimation considering only 1D spread. The equivalent delay spread can then be derived as
\begin{equation}
	\label{equivalent_spread}
	\begin{aligned}
		\tau_{\max}^{\prime}&=\tau_{\max}+a_{\max}T_{0}-(-a_{\max}T_{0})\\
		&=\tau_{\max}+2a_{\max}T_{0}.
	\end{aligned}
\end{equation}
Therefore, the potential diversity order $G$ can be approximately evaluated as $G=\frac{\tau_{\max}^{\prime}}{t_{r}}$. On the other hand, more received subcarriers are required to capture the power leakage to promote reliability. Let $Q_{1}=Q+\left\lceil\frac{a_{\max}T_{0}}{t_{r}}\right\rceil$ and $Q_{2}=Q+\left\lceil\frac{\tau_{\max}+a_{\max}T_{0}}{t_{r}}\right\rceil$ denote the leakage range for negative and positive equivalent delay, where $Q$ is the extra spread considering the power leakage brought by the off-grid effects in sinc function. It is very similar to the conventional zero padding in the frequency domain for UWA-OFDM systems \cite{UWA_OFDM_begin}. As a result, there are $N=M+Q_{1}+Q_{2}$ received filters with $n=-Q_{1},-Q_{1}+1,\cdots,M+Q_{2}-1$, where we have $t_{n}=t_{0}+nt_{r}$. Based on the above analysis, the equivalent channel matrix in \eqref{matrix_IO_1} can be represented as $\mathbf{H}_{nm}=\sum_{i=1}^{P}h_{i}q_{nm}^{i}$. Considering the equivalent delay spread as shown in \eqref{equivalent_spread}, $G=\frac{\tau_{\max}^{\prime}}{t_{r}}$ can also approximately measure the bandwidth of $\mathbf{H}$. To be more specific, Fig. \ref{Fig_eqChannel} illustrates an example of the equivalent channel matrix under simulation parameters. It is obvious that several resolvable paths can be observed with compact bandwidth in the channel matrix, which provides the possibility of multipath diversity gain in wideband LTV channels with low equalization complexity. Meanwhile, the band-limited property of the equivalent channel, illustrated in Fig. \ref{Fig_eqChannel}, contributes to a reduction in equalization complexity. This property can be leveraged, for instance, by employing LU decomposition to simplify the conventional LMMSE equalization.\par 
\subsection{Parameter Selection Issues}
\label{subsec_parameter}
In this subsection, the parameter selection for HFMC modulation is investigated. At first, $|t_{n}|<\epsilon T_{0}$ must be satisfied for the approximate orthogonality condition in \textbf{Theorem \ref{th1_ideal_correlation}} and the integrated 1D equivalent delay spread in \textbf{Theorem \ref{th3_IO}}, where $\epsilon$ denotes the error threshold adopted in prior analysis. Moreover, the equivalent delay range for transmit subcarriers is derived in \eqref{tm_range} under spectrum constraints. Considering the maximum equivalent delay spread introduced by the wideband LTV channels, as characterized in \textbf{Theorem \ref{th3_IO}} and Eq. \eqref{equivalent_spread}, this condition can be reformulated as follows
\begin{equation}
	\label{range_parameter}
	\begin{cases}
		T_{0}-T_{p}-\frac{K}{f_{c}+\frac{B}{2}}+\tau_{\max}+a_{\max}T_{0}<\epsilon{T_{0}}\\
		T_{0}+(1+a_{\max})T-\frac{K}{f_{c}-\frac{B}{2}}-a_{\max}T_{0}>-\epsilon{T_{0}}
	\end{cases},
\end{equation}
which serves as the basis for the parameter optimization. Since the FM rate $K$ and reference time $T_{0}$ are actually the basic parameters for HFMC modulation, we rewrite \eqref{range_parameter} as follows to see the relation between them more clearly
\begin{equation}
	\label{parameter_range_K}
	\begin{cases}
		K>(f_{c}+\frac{B}{2})(1+a_{\max}-\epsilon)T_{0}\\
		K<(f_{c}-\frac{B}{2})\Big((1-a_{\max}+\epsilon)T_{0}+(1+a_{\max})T\Big)
	\end{cases},
\end{equation}
where the first inequality is simplified by considering $T_{p}>\tau_{\max}$. However, the system constraints in \eqref{parameter_range_K} involve both $K$ and $T_{0}$, which hinders the determination of their respective allowable ranges. In order to proceed further in a tractable manner, we first analyze the feasible range of $T_{0}$ based on the above analysis, and then determine the value of $K$. To guarantee there are valid values for $K$ to satisfy \eqref{parameter_range_K}, we can derive
\begin{equation}
	\label{T0_range_max}
	c_{T_{0}}<\frac{(f_{c}-\frac{B}{2})(1+a_{\max})}{B+2f_{c}(a_{\max}-\epsilon)}.
\end{equation}
For notational simplicity, we define $c_{T_{0}}=\frac{T_{0}}{T}$ in \eqref{T0_range_max}. On the other hand, another inequality for $K$ should be considered as shown in \eqref{K_eq_1} to satisfy the spectrum constraints, which leads to the following formula for $T_{0}$ as
\begin{equation}
	\label{T0_range_min}
	\begin{aligned}
		c_{T_{0}}&>\frac{1}{1-a_{\max}+\epsilon}\\
		&\times\left(\frac{f_{c}-\frac{B}{2}}{B}(1+a_{\max})+\frac{f_{c}+\frac{B}{2}}{B}\frac{T_{p}}{T}\right).
	\end{aligned}
\end{equation}\par
However, the relations in \eqref{T0_range_max} and \eqref{T0_range_min} necessitate specifying the threshold $\epsilon$, whose admissible range remains unclear. To enable further tractable analysis, we proceed to determine the feasible range for $\epsilon$. From \eqref{T0_range_min}, $\frac{T_{0}}{T}$ must be positive since $a_{\max}=\frac{v_{\max}}{c}<1$ can be guaranteed for almost all communication objects. As a result, the right side in \eqref{T0_range_max} should be positive, which leads to
\begin{equation}
	\label{epsilon_max}
	\epsilon\leq a_{\max}+\frac{B}{2f_{c}}.
\end{equation}
Then similar to the derivations for \eqref{T0_range_min} and \eqref{T0_range_max}, there should be valid values for $\frac{T_{0}}{T}$ to satisfy these two inequalities. Therefore, we can derive
\begin{equation}
	\label{epsilon_min}
	\epsilon\geq\frac{c_{2}+\frac{2a_{\max}c_{2}f_{c}}{B}+c_{1}a_{\max}-c_{1}}{c_{1}+\frac{2c_{2}f_{c}}{B}},
\end{equation}
where we have
\begin{equation}
	\begin{cases}
		c_{1}&=\frac{(f_{c}-\frac{B}{2})(1+a_{\max})}{B}\\
		c_{2}&=c_{1}+\frac{f_{c}+\frac{B}{2}}{B}\frac{T_{p}}{T}
	\end{cases}.
\end{equation}
It is obvious that \eqref{epsilon_max} and \eqref{epsilon_min} do not include parameters to be determined, which can be employed to select appropriate error threshold value. Before the final deduction, the following theorem can be established to illustrate the basic trade-off between the spectral efficiency and potential diversity gain, which can be measured by the bandwidth of the equivalent channel matrix as $G=\frac{\tau_{\max}^{\prime}}{t_{r}}$.
\begin{theorem}
	\label{th4_para}
	\rm
	If $\epsilon$ satisfies \eqref{epsilon_max} and \eqref{epsilon_min}, the maximum number of transmit subcarriers $M$ in \eqref{system_capacity} increases while the maximum bandwidth $G=\frac{\tau_{\max}^{\prime}}{t_{r}}$ of the equivalent channel matrix decreases with $\frac{T_{0}}{T}$ increasing. When $T_{0}$ has been determined, both $G$ and $M$ enjoy the maximum value when $K$ is set as the maximum boundary in \eqref{parameter_range_K}, i.e.,
	\begin{equation}
		\label{K_value}
		K=\Big(f_{c}-\frac{B}{2}\Big)\Big((1-a_{\max}+\epsilon)T_{0}+(1+a_{\max})T\Big).
	\end{equation}
	\begin{IEEEproof}
		The proof is completed in Appendix \ref{th4_para_proof}.
	\end{IEEEproof}
\end{theorem}\par 
\begin{algorithm}
	\renewcommand{\algorithmicrequire}{\textbf{Input:}}
	\renewcommand{\algorithmicensure}{\textbf{Output:}}
	\caption{Parameter Selection for the Proposed HFMC Modulation}
	\label{alg_parameter}
	\begin{algorithmic}[1]
		\REQUIRE
		$f_{c}$, $B$, $T$, $T_{p}$, $a_{\max}$, $M_{s}$, $\Delta\epsilon$, and $\Delta{c}_{T_{0}}$
		\ENSURE
		$\epsilon$, $T_{0}$ and $K$
		\STATE 
		\textit{\% Determine $\epsilon$}
		\STATE 
		Compute $\epsilon\in(\epsilon_{\min},\epsilon_{\max})$ according to \eqref{epsilon_max} and \eqref{epsilon_min}
		\STATE 
		Initialize $\epsilon=\epsilon_{\min}$ and $M=0$
		\WHILE{$M<M_{s}$}
		\STATE 
		$\epsilon\leftarrow\epsilon+\Delta\epsilon$
		\STATE 
		Compute $c_{T_{0}}$ and $T_{0}=T\times c_{T_{0}}$ according to \eqref{T0_range_max}
		\STATE 
		Compute $K$ according to \eqref{K_value}
		\STATE 
		Compute $M$ according to \eqref{system_capacity}
		\ENDWHILE
		\STATE
		\textit{\% Determine $T_{0}$ and $K$}
		\STATE
		Initialize $c_{T_{0}}$ and $T_{0}=T\times c_{T_{0}}$ according to \eqref{T0_range_min}
		\STATE 
		Compute $K$ according to \eqref{K_value}
		\STATE 
		Compute $M$ according to \eqref{system_capacity}
		\WHILE{$M<M_{s}$}
		\STATE
		$c_{T_{0}}\leftarrow c_{T_{0}}+\Delta{c}_{T_{0}}$
		\STATE 
		$T_{0}=T\times c_{T_{0}}$
		\STATE 
		Compute $K$ according to \eqref{K_value}
		\STATE 
		Compute $M$ according to \eqref{system_capacity}
		\ENDWHILE
		\STATE
		Return $\epsilon$, $T_{0}$ and $K$
	\end{algorithmic}		
\end{algorithm} 
Based on \textbf{Theorem \ref{th4_para}} and the aforementioned range analysis of $\epsilon$ and $c_{T_{0}}$, various parameter selection strategies can be adopted. We establish a minimum number of subcarriers, $M_{s}$, as a design criterion to balance spectral efficiency and potential diversity gain, requiring that $M_{s}$ be less than the subcarrier count obtained when both $\epsilon$ and $T_{0}$ are maximized. To meet this criterion, the smallest $\epsilon$ that guarantees $M\geq{M_{s}}$ under maximum $T_{0}$ is selected, after which $T_{0}$ is minimized subject to $M=M_{s}$. The comprehensive selection procedure is illustrated in \textbf{Algorithm \ref{alg_parameter}}, where $\Delta\epsilon$ and $\Delta c_{T_{0}}$ represent the respective step sizes for updating $\epsilon$ and $c_{T_{0}}$. The algorithm commences with $\epsilon$ initialized to its minimum value as specified in \eqref{epsilon_min}. At each iterative step, $\epsilon$ is augmented by $\Delta\epsilon$, and the maximum number of subcarriers is computed based on $T_{0}$ and $K$ derived from \eqref{T0_range_max} and \eqref{K_value}. The iteration proceeds until the condition $M\geq{M_{s}}$ is met, thereby establishing the value of $\epsilon$. Subsequently, $c_{T_{0}}$ is initialized to its minimum value in accordance with \eqref{T0_range_min}. During each iteration, $c_{T_{0}}$ is increased by $\Delta{c_{T_{0}}}$, and $K$ is determined via \eqref{K_value} to compute the maximum number of transmit subcarriers according to \eqref{system_capacity}. The process is terminated once $M=M_{s}$ is achieved, which yields the values for $T_{0}$ and $K$, and completes the parameter selection flow.\par 
\section{Numerical Results}
\label{sec_simu}
\begin{table}
	\caption{Numerical Parameters}
	\centering
	\label{table_simu}
	\renewcommand\arraystretch{1.5}
	\begin{tabular}{p{17em}|p{10em}}
		\hline\hline
		Parameter &
		Typical value\\
		\hline
		Carrier frequency ($f{c}$) & 50 kHz\\
		\hline
		Bandwidth ($B$) & 5 kHz\\
		\hline
		Sampling frequency ($F_{s}$)& 40 kHz\\
		\hline
		Multicarrier symbol period ($T$) & 102.4 ms\\
		\hline
		Prefix Length ($T_{p}$) & 25.6 ms\\
		\hline
		Error threshold for HFMC ($\epsilon$) & 0.034\\
		\hline
		The number of subcarriers & 256\\
		\hline
		Mapping alphabet & QPSK, 16QAM\\
		\hline
		The number of paths ($P$) & 15\\
		\hline
		Maximum Doppler scaling factor ($a_{\max}$) & $3.43\times10^{-3}$\\
		\hline\hline
	\end{tabular}
\end{table}
In this section, numerical results are illustrated to confirm the performance of the proposed HFMC modulation in wideband LTV channels with both delay and Doppler scaling spread. Relevant parameters are presented in Table \ref{table_simu}, where the digital implementation is utilized for all modulation schemes by oversampling the continuous-time waveform 8 times. Following the parameter selection procedure outlined in \textbf{Algorithm \ref{alg_parameter}} with the step sizes as $\Delta\epsilon=10^{-3}$ and $\Delta{c}_{T_{0}}=10^{-2}$, the error threshold $\epsilon$ and the ratio $c_{T_{0}}$ are obtained as 0.034 and 24.52, respectively. Consequently, $T_{0}$ is computed as 2.511 s, while the equivalent delay $t_{m}$ lies within the range of $(-0.077, 0.051)$ s. These results validate the approximations in \textbf{Theorem \ref{th1_ideal_correlation}} and \textbf{Theorem \ref{th3_IO}} again, since the condition $|t_{m}|\ll T_{0}$ is satisfied. The maximum Doppler scaling factor is set as $3.43\times10^{-3}$, which caters to the maximum relative speed of $v_{\max}=10$ kn in UWA communications with $c\approx1500$ m/s. For each channel realization, the inter-arrival times are distributed exponentially with mean $\mathbb{E}\left(\tau_{i+1}-\tau_{i}\right)=1$ ms like \cite{UWA_OFDM_dl_samek,TSP_zhou_channel_simu_1}, which leads to the average delay spread of about $15$ ms with 15 incident paths. The path amplitudes are Rayleigh distributed with the average power decreasing exponentially with delay, where a total decay of 20 dB between the beginning and the end of the guard time of 25.6 ms is considered \cite{UWA_OFDM_dl_samek,TSP_zhou_channel_simu_1,OTFS_DSE_TVT}. The Doppler scaling factor of each path is randomly generated as $a_{i}=a_{\max}\cos(\theta_{i})$ with $\theta_{i}\sim\mathcal{U}(-\pi,\pi)$. We select OFDM, single-carrier, and ODDM modulation as the comparison baseline, where ODDM has been widely regarded as the candidate waveform for high-mobility scenarios \cite{ODDM_tutorial,ODDM_wideband_GC}. However, as a conventional DD domain modulation scheme, it suffers performance degradation in wideband LTV channels considering the frequency-dependent Doppler shift \cite{OTFS_DSE_TVT,OTFS_DSE_TWC}. To guarantee a fair comparison, all waveforms share the same spectral efficiency and parameters as in Table \ref{table_simu}, where the subcarrier spacing of OFDM can be derived as $\Delta{f}=19.53$ Hz while the implementation of ODDM modulation is approximated by the inverse discrete Zak transform (IDZT)-precoded single-carrier as illustrated in \cite{ODDM_tutorial,ODDM_IO_Tcom_24}. LMMSE equalization\footnote{The LMMSE equalizer is adopted in this work for performance validation since it is one of the most widely used detectors and serves as a standard benchmark \cite{OTFS_DSE_TWC,ref_AFDM,ref_LTVchannel_book,ODDM_IO_Tcom_24,ref_MMSE_evaluation}. Its low-complexity implementation leverages the band-limited structure of the equivalent channel matrix in the proposed HFMC modulation, achieving linear complexity \cite{ref_LMMSE_Lowcom}. It is worth noting that other detection algorithms, such as message passing \cite{ref_OTFS} and orthogonal approximate message passing \cite{ref_OAMP}, can be directly applied to the proposed waveform based on its matrix-based formulation in \eqref{matrix_IO_1}, without requiring any modifications. Further exploration of these advanced detectors tailored to the proposed modulation is a promising direction for future research.}, given its ubiquity in modern communication systems, is adopted for all modulation schemes, where the inter-carrier interference is fully considered for OFDM schemes to ensure reliability and exploit the potential Doppler resolution. \par 
\begin{figure}
	\centering{\includegraphics[width=1\linewidth]{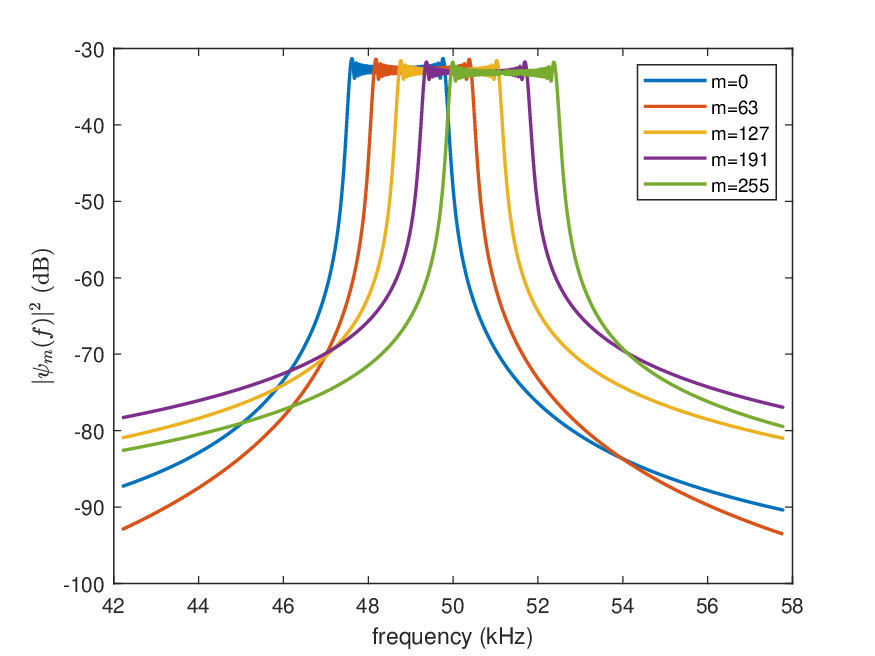}}
	\caption{Spectrum of selected subcarriers.}
	\label{Fig_spectrum}	
\end{figure}
Fig. \ref{Fig_spectrum} presents the spectrum of selected transmit subcarriers under the parameters in Table \ref{table_simu}. It verifies the correctness of \textbf{Theorem \ref{th2_spectrum}} since the spectrum can be approximately confined to the designated frequency band. On the other hand, an equivalent frequency shift can be observed among subcarriers, since each subcarrier is a variant of the initial HFM signal. The spectrum of each subcarrier is almost flat within its band, which aligns with the analysis in \eqref{spectrum_subcarrier}. Notably, the subcarriers exhibit significant spectral overlap. Although the term ``subcarrier'' is retained for analogy, these waveforms are not Fourier bases. Instead, they maintain approximate orthogonality via matched filtering as established in \textbf{Theorem \ref{th1_ideal_correlation}}. This overlap arises from the HFM structure and does not compromise orthogonality because the frequency-domain separation is not required. In fact, such overlap is beneficial for improving spectral efficiency, since it allows more subcarriers to be packed within the available bandwidth. This overlap also allows each subcarrier to occupy a wider bandwidth, which is beneficial for combating multipath fading in wideband LTV channels compared with OFDM systems.\par 
\begin{figure}
	\centering{\includegraphics[width=1\linewidth]{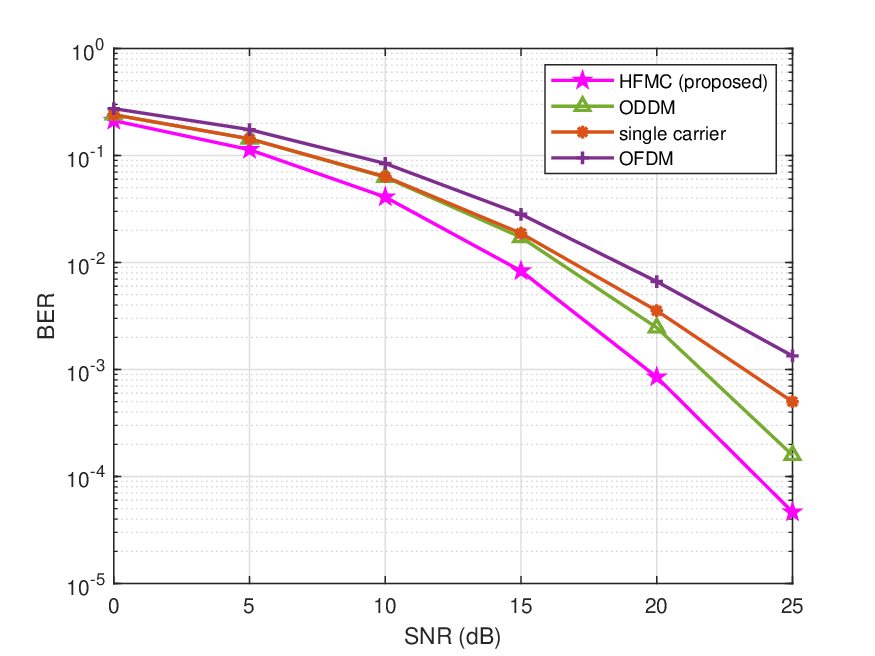}}
	\caption{BER against SNR under QPSK alphabets.}
	\label{Fig_BER_SNR_QPSK}	
\end{figure}
Fig. \ref{Fig_BER_SNR_QPSK} depicts the BER performance against the signal-to-noise ratio (SNR) under QPSK alphabets. It is obvious that the proposed HFMC modulation reveals the lowest BER in wideband LTV channels. When SNR=20 dB, the BER of the proposed HFMC modulation is lower than $10^{-3}$, which is about half of that in ODDM systems. When BER is about $10^{-3}$, the proposed HFMC modulation outperforms the ODDM systems with more than $1$ dB. The reason behind it is that ODDM is designed mainly for narrowband doubly dispersive channels, where the high mobility can be approximated with a constant Doppler shift across the bandwidth. Therefore, it is not optimal under wideband LTV channels with Doppler scaling spread. However, since the channel reveals the doubly selective property, ODDM still outperforms conventional OFDM and single carrier systems due to the efficient spread in the time-frequency domain.\par 
\begin{figure}
	\centering{\includegraphics[width=1\linewidth]{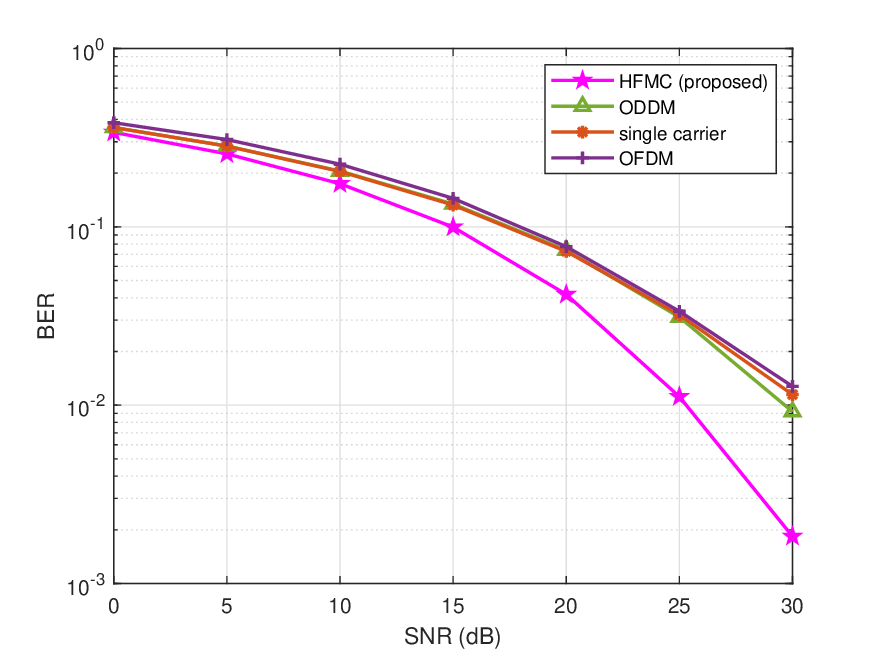}}
	\caption{BER against SNR under 16QAM alphabets.}
	\label{Fig_BER_SNR_16QAM}	
\end{figure}
In Fig. \ref{Fig_BER_SNR_16QAM}, we illustrate the performance by plotting BER against SNR under 16QAM alphabets. The performance precedence is amplified due to the higher order of alphabets, where the SNR gap between HFMC and other modulation schemes is about 5 dB when BER is $10^{-2}$. The performance superiority can also be embodied by the slope, which demonstrates that the proposed HFMC modulation can better exploit the multipath diversity in wideband LTV channels. Finally, the mismatch between the Doppler scaling effect and conventional shift assumptions makes ODDM systems suffer more severe degradation, where similar BER performance can be observed for ODDM, OFDM, and single-carrier systems even though in high SNR scenarios, e.g., SNR=30 dB.\par 
\begin{figure}
	\centering{\includegraphics[width=1\linewidth]{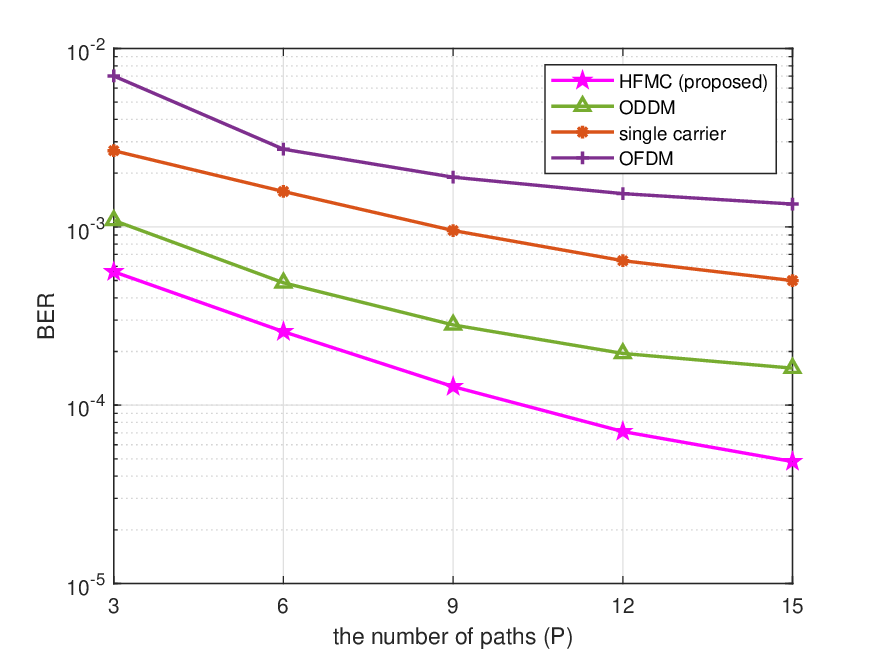}}
	\caption{BER against $P$ under SNR=25 dB and QPSK alphabets.}
	\label{Fig_BER_P_QPSK_25dB}	
\end{figure} 
The performance in wideband LTV channels is also demonstrated by plotting BER against the number of paths $P$ in Fig. \ref{Fig_BER_P_QPSK_25dB} under SNR=25 dB and QPSK alphabets. It is apparent that all modulation schemes enjoy enhanced reliability with the number of paths increasing thanks to better multipath diversities. The performance advantage of the proposed HFMC modulation is also amplified with the number of paths. For example, when there are $12$ incident paths, the BER of the proposed HFMC modulation is about $7\times10^{-5}$ while that of the ODDM system is still higher than $10^{-4}$. When $P$ is increased to $15$, only a slight BER decrease occurs for ODDM while the BER of the proposed HFMC schemes could be reduced to lower than $5\times10^{-5}$. On the other hand, the BER of OFDM is still higher than $10^{-3}$ even though the inter-carrier interference has been taken into consideration to carry out the LMMSE equalization. It also indicates the necessity to develop novel multicarrier frameworks for high-mobility applications. Based on the BER evaluation, we deduce that the proposed HFMC modulation can serve as a qualified candidate to implement multicarrier communications in wideband LTV channels.\par 
\begin{figure}
	\centering{\includegraphics[width=1\linewidth]{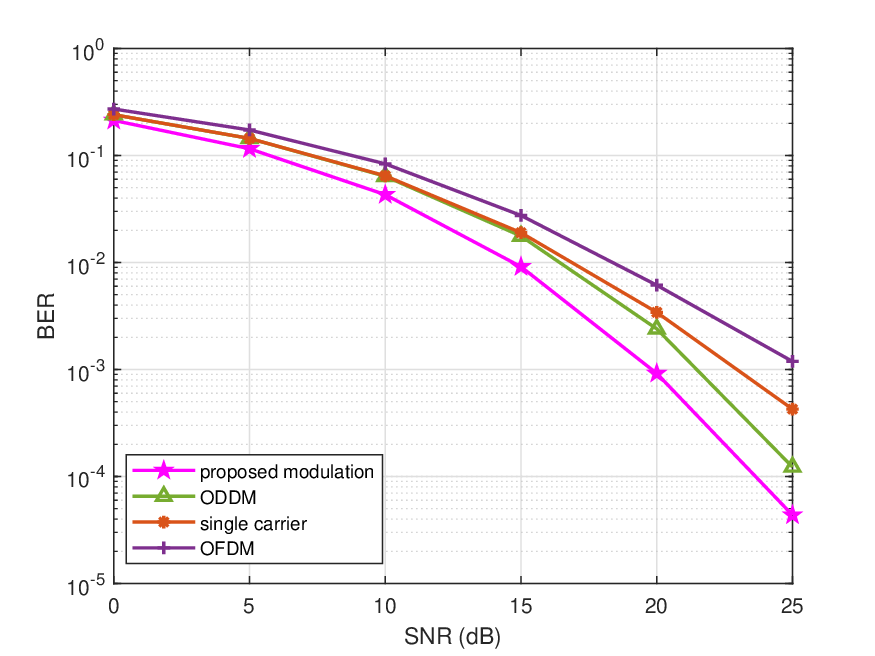}}
	\caption{BER against SNR under QPSK alphabets and 512 transmit subcarriers.}
	\label{Fig_BER_M512}	
\end{figure}
Fig. \ref{Fig_BER_M512} illustrates the BER against SNR under 512 subcarriers and QPSK alphabets, confirming the superiority of the proposed HFMC modulation under different numerologies. The proposed scheme achieves the lowest BER across the entire SNR range. When SNR is 25 dB, the BER of HFMC is approximately $4\times10^{-5}$, whereas ODDM systems still exhibit a BER above $10^{-4}$, revealing a reduction of more than 50\%. For a target BER of $10^{-4}$, the proposed HFMC modulation outperforms the other waveforms by more than 1.6 dB. These results demonstrate that HFMC is a promising candidate waveform for reliable transmission in wideband LTV channels under diverse parameter settings.\par 
\begin{figure}
	\centering{\includegraphics[width=1\linewidth]{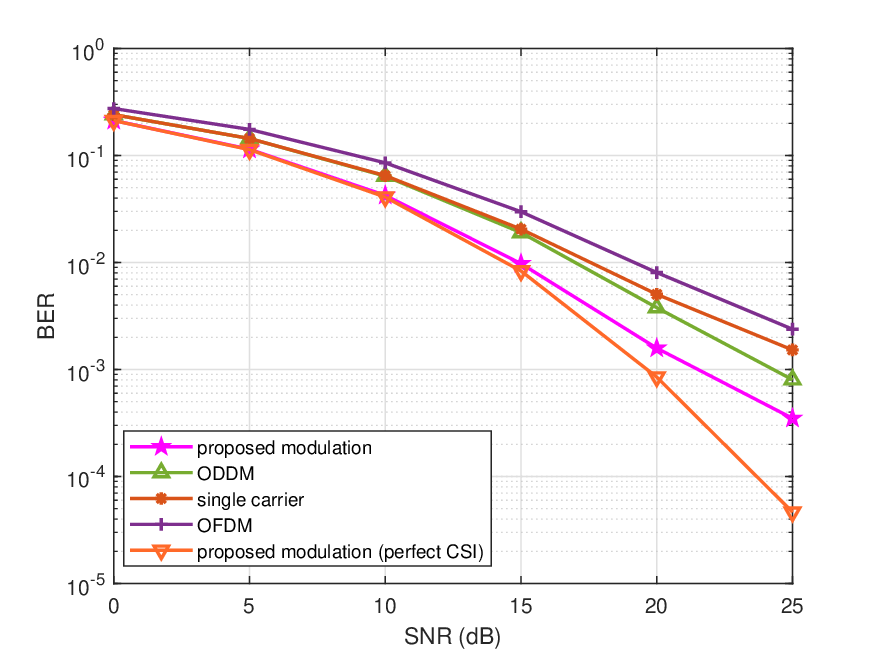}}
	\caption{BER against SNR under QPSK alphabets and imperfect CSI.}
	\label{Fig_BER_imperfect}	
\end{figure} 
Finally, Fig. \ref{Fig_BER_imperfect} evaluates the robustness by plotting BER against SNR under QPSK alphabets and imperfect channel state information (CSI). Following \cite{Simu_ICSI,OTFS_DSE_TVT}, the gain of each path is assumed to be corrupted by additive Gaussian noise, leading to a normalized MSE (NMSE) of 1\%. As observed, the proposed modulation consistently achieves the lowest BER among all considered schemes, even considering the effect of imperfect CSI. For instance, at the BER of $10^{-3}$, the SNR penalty incurred by the proposed modulation due to imperfect CSI is only less than $2$ dB relative to the ideal case with perfect CSI. In contrast, the next-best alternative, ODDM, requires approximately $5$ dB higher SNR to attain the same BER level. These results verify the robustness of the proposed HFMC modulation against channel estimation errors, making it a promising candidate for practical communications under wideband LTV channels.\par 
\section{Conclusion}
\label{sec_conclusion}
Inspired by the 1D compact representation of discrete-time doubly dispersive channels and the potential diversity gain of AFDM systems, we proposed the HFMC modulation for wideband LTV channels in this paper, where the passband continuous-time representation was utilized. By adopting HFM signals, the Doppler scaling factor introduced by the high-mobility channel could be absorbed into an equivalent time delay. It motivated the principle of HFMC modulation by investigating the approximate orthogonality among subcarriers, which were generated from a basis HFM signal by utilizing uniformly spaced equivalent time delay. The spectrum of HFMC subcarriers was also analyzed to measure the capacity and establish the basis for parameter selection. The input-output relation was then characterized, which demonstrated that the 2D multipath spread can be approximately integrated into a 1D equivalent delay spread. The parameter selection framework was finally developed to achieve the flexible trade-off between the spectral efficiency and maximum potential diversity order. Finally, numerical results confirm the excellent reliability of the proposed HFMC modulation schemes. For future work, it is meaningful to execute deeper research in relevant channel estimation and data detection issues for practical deployment.\par 
\appendices
\section{Proof of Theorem \ref{th1_ideal_correlation}}
\label{proof_th1_ideal_correlator}
We first compute the value for $A_{m}$ to normalize the power for the $m$-th subcarrier. Based on \eqref{subcarrier_form}, we can derive
\begin{equation}
	\begin{aligned}
		&\int_{0}^{T}\varphi_{m}^{*}(t)\varphi_{m}(t)dt=A_{m}^{2}\int_{0}^{T}\frac{1}{\left(1+\frac{t-t_{m}}{T_{0}}\right)^{2}}dt\\
		&\overset{(a)}{=}A_{m}^{2}T_{0}\int_{u(0)}^{u(T)}\frac{1}{(1+u)^{2}}du\\
		&=\frac{A_{m}^{2}T_{0}^{2}T}{(T_{0}-t_{m})(T_{0}+T-t_{m})},	
	\end{aligned}
\end{equation}
where we have $u(t)=\frac{t-t_{m}}{T_{0}}$ in $(a)$. As a result, to normalize the power for each subcarrier, $A_{m}$ can be set as follows
\begin{equation}
	\label{proof_Am}
	A_{m}=\sqrt{\frac{(T_{0}+T-t_{m})(T_{0}-t_{m})}{T_{0}^{2}T}}.
\end{equation}\par 
\begin{figure*}
	\begin{equation}
		\label{correlation_proof_eq}
		\begin{aligned}
			\int_{0}^{T}\varphi_{n}^{*}(t)\varphi_{m}(t)&=A_{m}A_{n}\int_{0}^{T}\frac{1}{\left(1+\frac{t-t_{n}}{T_{0}}\right)\left(1+\frac{t-t_{m}}{T_{0}}\right)}e^{j2\pi{K}\bigg(\ln\left(1+\frac{t-t_{m}}{T_{0}}\right)-\ln\left(1+\frac{t-t_{n}}{T_{0}}\right)\bigg)}~dt\\
			&\overset{(a)}{=}\frac{A_{m}A_{n}T_{0}^{2}}{t_{m}-t_{n}}\int_{u(0)}^{u(T)}e^{j2\pi{K}u}~du\\
			&=\frac{A_{m}A_{n}T_{0}^{2}}{\pi{K}(t_{m}-t_{n})}e^{j\pi{K}(u(0)+u(T))}\sin\bigg(\pi{K}\Big(u(T)-u(0)\Big)\bigg)\\
			&\overset{(b)}{=}\frac{A_{m}A_{n}T_{0}^{2}T}{(T_{0}+T-t_{n})(T_{0}-t_{m})}e^{j\pi{K}\theta_{nm}}\frac{\sin\pi{K}\ln\left(1+\frac{(t_{m}-t_{n})T}{(T_{0}+T-t_{n})(T_{0}-t_{m})}\right)}{\pi{K}\frac{(t_{m}-t_{n})T}{(T_{0}+T-t_{n})(T_{0}-t_{m})}}\\
			&\overset{(c)}{\approx}\frac{A_{m}A_{n}T_{0}^{2}T}{(T_{0}+T-t_{n})(T_{0}-t_{m})}e^{j\pi{K}\theta_{nm}}\frac{\sin\pi{K}\left(\frac{(t_{n}-t_{m})T}{(T_{0}+T-t_{n})(T_{0}-t_{m})}\right)}{\pi{K}\frac{(t_{n}-t_{m})T}{(T_{0}+T-t_{n})(T_{0}-t_{m})}}\\
			&\overset{(d)}{\approx}e^{j\pi{K}\theta_{nm}}\frac{\sin\pi{K}\left(\frac{(t_{n}-t_{m})T}{(T_{0}+T)T_{0}}\right)}{\pi{K}\frac{(t_{n}-t_{m})T}{(T_{0}+T)T_{0}}}
		\end{aligned}
	\end{equation}
	\hrulefill
\end{figure*}
Based on $A_{m}$ in \eqref{proof_Am}, \eqref{correlation_proof_eq} can be derived at the top of the next page, where $u(t)$ in $(a)$ can be derived as
\begin{equation}
	u(t)=\ln\left(1+\frac{t-t_{m}}{T_{0}}\right)-\ln\left(1+\frac{t-t_{n}}{T_{0}}\right).
\end{equation}
$\theta_{nm}$ in $(b)$ can be derived as
\begin{equation}
	\begin{aligned}
		\theta_{nm}&=u(0)+u(T)\\
		&=\ln\left(\frac{(T_{0}-t_{m})(T_{0}+T-t_{m})}{(T_{0}-t_{n})(T_{0}+T-t_{n})}\right).
	\end{aligned}
\end{equation}
The approximation in $(c)$ requires $\frac{(t_{m}-t_{n})T}{(T_{0}+T-t_{n})(T_{0}-t_{m})}\ll1$ while the approximation in $(d)$ requires $|t_{m}|\ll{T_{0}}$ for all subcarriers, which can also strengthen the approximation in $(c)$. After the derivations in \eqref{correlation_proof_eq}, the proof of basic conclusions in \textbf{Theorem \ref{th1_ideal_correlation}} is completed.\par 
\section{Proof of Theorem \ref{th2_spectrum}}
\label{proof_th2_spectrum}
It is almost impossible to compute the precise spectrum for HFM signals considering the complicated form of phase variation. However, we can provide an approximate evaluation based on the method of stationary phase \cite{stationry_phase_method}, whose approximation precision can be verified by the large product between the time duration and bandwidth. At first, the instantaneous frequency for $\varphi_{m}(t)$ can be derived as
\begin{equation}
	\label{frequency_instantaneous}
	f_{m}(t)=\frac{K}{T_{0}+t-t_{m}}.
\end{equation}
To ensure precision, the product between the time duration and bandwidth should be much larger than $1$. For the $m$-th transmit subcarrier, the time duration is larger than $T$. For the bandwidth, we have
\begin{equation}
	B_{m}>f_{m}(T)-f_{m}(0)=\frac{KT}{(T_{0}+T-t_{m})(T_{0}-t_{m})}.
\end{equation}
Therefore, the product between the time duration and bandwidth can be approximately measured by 
\begin{equation}
	B_{m}T>\frac{KT^{2}}{T_{0}(T_{0}+T)},
\end{equation}
where we assume $|t_{m}|\ll{T_{0}}$. As a result, the method of stationary phase can be utilized to evaluate the band occupation if $KT^{2}\gg T_{0}(T_{0}+T)$. Based on the analysis in \cite{stationry_phase_method}, the spectrum range can be computed as $f\in\left(f_{m}\Big((1+a_{\max})T\Big),f_{m}(-T_{p})\right)$, where we have
\begin{equation}
	\begin{cases}
		f_{m}\Big((1+a_{\max})T\Big)=\frac{K}{T_{0}+(1+a_{\max})T-t_{m}}\\
		f_{m}(-T_{p})=\frac{K}{T_{0}-T_{p}-t_{m}}\\
	\end{cases}.
\end{equation}
On the other hand, the first-order derivative of $f_{m}(t)$ can be derived as
\begin{equation}
	\label{derivative_frequency}
	f_{m}^{\prime}(t)=\frac{-K}{(T_{0}+t-t_{m})^{2}}.
\end{equation}\par
Now let $\psi_{m}(f)$ denote the spectrum of $\varphi_{m}(t)$. Based on the aforementioned analysis, for $f\notin\left(f_{m}\Big((1+a_{\max})T\Big),f_{m}(-T_{p})\right)$, we can approximately derive $\psi_{m}(f)\approx0$. For $f\in\left(f_{m}\Big((1+a_{\max})T\Big),f_{m}(-T_{p})\right)$, the corresponding time to have the instantaneous frequency $f$ can be derived as
\begin{equation}
	t_{f}=\frac{K}{f}+t_{m}-T_{0}.
\end{equation}
Therefore, $\psi_{m}(f)$ can be approximately derived as \cite{stationry_phase_method}
\begin{equation}
	\begin{aligned}
		\psi_{m}(f)&=\int_{-T_{p}}^{(1+a_{\max})T}\varphi_{m}(t)e^{-j2\pi ft} dt\\
		&\approx\frac{A_{m}\sqrt{\frac{1}{f_{m}^{\prime}(t_{f})}}}{1+\frac{t_{f}-t_{m}}{T_{0}}}e^{j2\pi\Big(\ln\left(1+\frac{t_{f}-t_{m}}{T_{0}}\right)-ft_{f}\Big)}e^{-j\frac{\pi}{4}}\\
		&=\frac{A_{m}T_{0}e^{-j\frac{\pi}{4}}}{\sqrt{K}}e^{j2\pi\left(K\ln\left(\frac{K}{fT_{0}}\right)-K-f(t_{m}-T_{0})\right)},
	\end{aligned}
\end{equation}
which completes the proof of \textbf{Theorem \ref{th2_spectrum}}.\par 
\section{Proof of Theorem \ref{th3_IO}}
\label{proof_th3_IO}
At first, we have
\begin{equation}
	\begin{aligned}
		&y_{n}=\int_{0}^{T}\varphi_{n}^{*}(t)r(t)dt\\
		&=\int_{0}^{T}\varphi_{n}^{*}(t)\left(\sum_{i=1}^{P}h_{i}s\Big((1+a_{i})t-\tau_{i}\Big)+w(t)\right)dt\\
		&=\int_{0}^{T}\varphi_{n}^{*}(t)\left(\sum_{i=1}^{P}h_{i}\sum_{m=0}^{M-1}x_{m}\varphi_{m}\Big((1+a_{i})t-\tau_{i}\Big)\right)dt+w_{n}\\
		&=\sum_{m=0}^{M-1}x_{m}\left(\sum_{i=1}^{P}h_{i}\int_{0}^{T}\varphi_{n}^{*}(t)\varphi_{m}\Big((1+a_{i})t-\tau_{i}\Big)\right)dt+w_{n}\\
		&=\sum_{m=0}^{M-1}\left(\sum_{i=1}^{P}h_{i}q_{nm}^{i}\right)x_{m}+w_{n},\\
	\end{aligned}
\end{equation}
where we have $q_{nm}^{i}=\int_{0}^{T}\varphi_{n}^{*}(t)\varphi_{m}\Big((1+a_{i})t-\tau_{i}\Big)dt$. Based on \eqref{eqdelay_double_channel}, we have $\varphi_{m}\Big((1+a_{i})t-\tau_{i}\Big)=A_{m}\frac{e^{j2\pi\ln(1+a_{i})}}{1+a_{i}}g(t-t_{m,i}^{\prime})$ with $t_{m,i}^{\prime}=\frac{\tau_{i}+t_{m}+a_{i}T_{0}}{1+a_{i}}$. Similar to \eqref{correlation_proof_eq}, we can derive
\begin{equation}
	\begin{aligned}
		q_{nm}^{i}&=\frac{A_{m}A_{n}T_{0}^{2}T}{(T_{0}+T-t_{n})(T_{0}-t_{m,i}^{\prime})}\times\frac{e^{j2\pi\ln(1+a_{i})}}{1+a_{i}}\\
		&\times e^{j\pi K\eta_{nm}^{i}}\frac{\sin{\pi K\ln\left(1+\frac{(t_{n}-t_{m,i}^{\prime})T}{(T_{0}+T-t_{n})(T_{0}-t_{m,i}^{\prime})}\right)}}{\pi K\frac{(t_{n}-t_{m,i}^{\prime})T}{(T_{0}+T-t_{n})(T_{0}-t_{m,i}^{\prime})}}\\
		&\overset{(a)}{\approx}\frac{A_{m}A_{n}T_{0}^{2}T}{(T_{0}+T-t_{n})(T_{0}-t_{m,i}^{\prime})}\times\frac{e^{j2\pi\ln(1+a_{i})}}{1+a_{i}}\\
		&\times e^{j\pi K\eta_{nm}^{i}} \text{sinc}\left(\frac{K(t_{n}-t_{m,i}^{\prime})T}{(T_{0}+T-t_{n})(T_{0}-t_{m,i}^{\prime})}\right),
	\end{aligned}
\end{equation}
where we have 
\begin{equation}
	\eta_{nm}^{i}=\ln\left(\frac{(T_{0}-t_{m,i}^{\prime})(T_{0}+T-t_{m,i}^{\prime})}{(T_{0}-t_{n})(T_{0}+T-t_{n})}\right).
\end{equation}
The approximation in $(a)$ utilizes similar conditions to \textbf{Theorem \ref{th1_ideal_correlation}} to guarantee the linearization of logarithmic terms.\par 
\begin{figure*}
	\begin{equation}
		\label{proof_th4_M}
		\begin{aligned}
			c-d&=(f_{c}-\frac{B}{2})T(1-a_{\max}+\epsilon)\left(\frac{(1-a_{\max}+\epsilon)B}{f_{c}+\frac{B}{2}}-\frac{2(1+a_{\max})B}{f_{c}+\frac{B}{2}}+\frac{T_{p}}{T}+(1+a_{\max})\right)\\
			&=(f_{c}-\frac{B}{2})T(1-a_{\max}+\epsilon)\left(\frac{T_{p}}{T}+\frac{(1+a_{\max})f_{c}-(\frac{1}{2}+\frac{5}{2}a_{\max}-\epsilon)B}{f_{c}+\frac{B}{2}}\right)\\
			&>(f_{c}-\frac{B}{2})T(1-a_{\max}+\epsilon)\left(\frac{f_{c}-(\frac{1}{2}+\frac{5}{2}a_{\max})B}{f_{c}+\frac{B}{2}}\right)\overset{(a)}{>}0
		\end{aligned}
	\end{equation}
	\hrulefill
\end{figure*}
The deduction can be further simplified by considering the relation between $t_{m,i}^{\prime}$ and $a_{i}$. At first, we have
\begin{equation}
	\begin{aligned}
		&(1+a_{i})(T_{0}-t_{m,i}^{\prime})\\
		&=(1+a_{i})T_{0}-(\tau_{i}+t_{m}+a_{i}T_{0})\\
		&=T_{0}-t_{m}-\tau_{i},
	\end{aligned}
\end{equation}
and 
\begin{equation}
	\begin{aligned}
		&(1+a_{i})(T_{0}+T-t_{m,i}^{\prime})\\
		&=(1+a_{i})T_{0}+(1+a_{i})T-(\tau_{i}+t_{m}+a_{i}T_{0})\\
		&=T_{0}+(1+a_{i})T-t_{m}-\tau_{i}.
	\end{aligned}
\end{equation}
Therefore, the phase $\theta_{nm}^{i}$ can be derived as
\begin{equation}
	\begin{aligned}
		&\theta_{nm}^{i}=\eta_{nm}^{i}+2\ln(1+a_{i})\\
		&=\ln\left(\frac{(T_{0}-t_{m}-\tau_{i})(T_{0}+(1+a_{i})T-t_{m}-\tau_{i})}{(T_{0}-t_{n})(T_{0}+T-t_{n})}\right).
	\end{aligned}
\end{equation}
On the other hand, we have
\begin{equation}
	\begin{aligned}
		&(1+a_{i})(t_{n}-t_{m,i}^{\prime})\\
		&=(1+a_{i})t_{n}-t_{m}-(\tau_{i}+a_{i}T_{0}).
	\end{aligned}	
\end{equation}
It leads to the overall simplification as
\begin{equation}
	\label{simplify_delayscale_proof}
	\begin{aligned}
		&q_{nm}^{i}=\frac{A_{m}A_{n}T_{0}^{2}Te^{j\pi K\theta_{nm}^{i}}}{(T_{0}+T-t_{n})(T_{0}-t_{m}-\tau_{i})}\\
		&\times  \text{sinc}\left(\frac{KT\big((1+a_{i})t_{n}-t_{m}-(\tau_{i}+a_{i}T_{0})\big)}{(T_{0}+T-t_{n})(T_{0}-t_{m}-\tau_{i})}\right)\\
		&\overset{(a)}{\approx}e^{j\pi K\theta_{nm}^{i}}\text{sinc}\left(\frac{KT\Big(t_{n}-t_{m}-(\tau_{i}+a_{i}T_{0})\Big)}{(T_{0}+T)T_{0}}\right),
	\end{aligned}
\end{equation}
where $(a)$ holds if we have $|t_{n}|, |t_{m}+\tau_{i}|\ll{T_{0}}$. The proof of \textbf{Theorem \ref{th3_IO}} is then completed.\par 
\section{Proof of Theorem \ref{th4_para}}
\label{th4_para_proof}
For ease of illustration, we adopt the notation $\lambda=\frac{T_{0}}{T}$ to represent $T_{0}$ since $T$ is a known parameter. The following lemma is first established to serve as the basis for proving \textbf{Theorem \ref{th4_para}}.
\begin{lemma}
	\label{lemma_proof_th4}
	\rm 
	Let $f(\lambda)=\frac{c\lambda^{2}+d\lambda+e}{\lambda(\lambda+1)}$ denote the function for $\lambda$ to be investigated, which is utilized for determining the maximum number of transmit subcarriers and bandwidth of the channel matrix. If we have $c>d$ and $e<0$, $f(\lambda)$ increases with $\lambda$ increasing for $\lambda\in(0,+\infty)$. If we have $c<d$ and $e>0$, $f(\lambda)$ decreases with $\lambda$ increasing for $\lambda\in(0,+\infty)$.
	\begin{IEEEproof}
		The derivative of $f(\lambda)$ can be derived as
		\begin{equation}
			f^{\prime}(\lambda)=\frac{1}{\lambda^{2}(\lambda+1)^{2}}\Big((c-d)\lambda^{2}-2e\lambda-e\Big).
		\end{equation}
		If we have $c>d$ and $e<0$, $f^{\prime}(\lambda)>0$ can be ensured for $\lambda\in(0,+\infty)$. Therefore, $f(\lambda)$ increases with $\lambda$ increasing. Similarly, If we have $c<d$ and $e>0$, $f^{\prime}(\lambda)<0$ can be obtained for $\lambda\in(0,\infty)$. As a result, $f(\lambda)$ decreases with $\lambda$ increasing. The proof of \textbf{Lemma \ref{lemma_proof_th4}} can then be completed to serve as the basis for the following analysis.
	\end{IEEEproof}
\end{lemma}
We first investigate the maximum number of subcarriers in \eqref{system_capacity}, which measures the system capacity. When $T_{0}$ has been decided, it is obvious that $M$ increases with $K$ increasing. Selecting $K$ as \eqref{K_value} can achieve the maximum spectral efficiency with a certain $T_{0}$. By replacing $K$ in \eqref{system_capacity} with \eqref{K_value}, \eqref{system_capacity} can be simplified as
\begin{equation}
	M=\left\lceil\frac{c\lambda^{2}+d\lambda+e}{\lambda(\lambda+1)}\right\rceil,
\end{equation}
where we have 
\begin{equation}
	\begin{cases}
		c&=\frac{f_{c}-\frac{B}{2}}{f_{c}+\frac{B}{2}}BT(1-a_{\max}+\epsilon)^{2}\\
		d&=(f_{c}-\frac{B}{2})(1-a_{\max}+\epsilon)\\
		&\times\left(\frac{2(1+a_{\max})BT}{f_{c}+\frac{B}{2}}-T_{p}-(1+a_{\max})T\right)\\
		e&=-(1+a_{\max})(f_{c}-\frac{B}{2})T_{p}\\
		&-(f_{c}-\frac{B}{2})T(1+a_{\max})^{2}\frac{f_{c}-\frac{B}{2}}{f_{c}+\frac{B}{2}}\\
	\end{cases}.
\end{equation}\par 
It is obvious that $e<0$ holds. On the other hand, we can derive \eqref{proof_th4_M} at the top of this page to verify $c>d$, where $(a)$ is deduced by considering $\frac{f_{c}}{B}>1>\frac{1}{2}+\frac{5a_{\max}}{2}$ for $a_{\max}<0.2$. Therefore, $M$ increases with $\lambda=\frac{T_{0}}{T}$ increasing by utilizing the first case of \textbf{Lemma \ref{lemma_proof_th4}}.\par 
Since the bandwidth $G$ of the equivalent channel matrix indicates the maximum potential diversity order, it is required to be as large as possible. Since we have $G=\frac{\tau_{\max}^{\prime}}{t_{r}}$, it reaches the maximum value when $K$ is selected as in \eqref{K_value} since the equivalent delay resolution $t_{r}=\frac{T_{0}(T_{0}+T)}{KT}$ decreases with $K$ increasing if $T_{0}$ has been determined. Therefore, $K$ is set as \eqref{K_value}. Then similar to the prior analysis, the bandwidth $G$ can be rewritten as
\begin{equation}
	\begin{aligned}
		G&=\frac{\tau_{\max}^{\prime}}{t_{r}}=\frac{KT(\tau_{\max}+2a_{\max}T_{0})}{T_{0}(T_{0}+T)}\\
		&=(f_{c}-\frac{B}{2})T\times\frac{c\lambda^{2}+d\lambda+e}{\lambda(\lambda+1)},
	\end{aligned}
\end{equation} 
where we have
\begin{equation}
	\begin{cases}
		c&=2a_{\max}(1-a_{\max}+\epsilon)\\
		d&=2a_{\max}(1+a_{\max})+(1-a_{\max}+\epsilon)\frac{\tau_{\max}}{T}\\
		e&=(1+a_{\max})\frac{\tau_{\max}}{T}\\
	\end{cases}.
\end{equation}
It is obvious that $e>0$. On the other hand, we can derive 
\begin{equation}
	\label{proof_bandwidth_increase_final}
	\begin{aligned}
		&c-d=2a_{\max}(\epsilon-2a_{\max})-(1-a_{\max}+\epsilon)\frac{\tau_{\max}}{T}\\
		&=2a_{\max}\epsilon-4a_{\max}^{2}-\frac{\tau_{\max}}{T}-\epsilon\frac{\tau_{\max}}{T}+a_{\max}\frac{\tau_{\max}}{T}\\
		&<\epsilon(2a_{\max}-\frac{\tau_{\max}}{T}-\frac{0.8\tau_{\max}}{\epsilon T})+(a_{\max}-0.2)\frac{\tau_{\max}}{T}.
	\end{aligned}
\end{equation}
\eqref{proof_bandwidth_increase_final} indicates that if $2a_{\max}\epsilon<\frac{0.8\tau_{\max}}{T}$, $c<d$ can be easily obtained considering $a_{\max}<0.2$. In fact, we tend to select $\epsilon<0.1$ to guarantee the approximation precision and $a_{\max}<0.01$ can be acquired in most applications even though extremely high mobility is considered. It means $c<d$ holds if $\frac{\tau_{\max}}{T}>2.5\times10^{-3}$, which satisfies most multicarrier settings \cite{ref_multicarrier,UWA_OFDM_dl_samek,TSP_zhou_channel_simu_1}. Therefore, $G$ decreases with $\lambda$ increasing, which completes the proof of \textbf{Theorem \ref{th4_para}}.\par 
\bibliographystyle{IEEEtran}
\bibliography{ref-sum}	
\end{document}